%% file: env_epi_main_revised_clean.tex
\documentclass[12pt, titlepage]{article}
\usepackage{fullpage,setspace}
\usepackage{amssymb, amsthm, amsmath}
\usepackage{graphicx}
\usepackage{bm}
\usepackage{lineno}
\usepackage{times}
\usepackage{algorithm,algorithmic}
\usepackage{hyperref}
\usepackage{endfloat}
\usepackage{multirow}
\usepackage{authblk}

\input{commands}

\providecommand{\keywords}[1]
{
  \small	
  \textbf{\textit{Keywords---}} #1
}
    \usepackage[natbib=true, autocite=superscript, sorting=none, maxbibnames=6, minbibnames=3]{biblatex}

\hypersetup{
    colorlinks=true,
    linkcolor=blue,
    filecolor=magenta,      
    urlcolor=cyan,
    pdftitle={Overleaf Example},
    pdfpagemode=FullScreen,
    citecolor=blue
    }

\title{When are novel methods for analyzing complex chemical mixtures in epidemiology beneficial?}

\author[1, *]{Nate Wiecha}
\author[1]{Emily Griffith}
\author[1]{Brian J. Reich}
\author[2]{Jane A. Hoppin}
\affil[1]{Department of Statistics, North Carolina State University, Raleigh, North Carolina, U.S.A}
\affil[2]{Department of Biological Sciences, North Carolina State University, Raleigh, North Carolina, U.S.A}
\affil[*]{Corresponding author, email: nbwiecha@ncsu.edu}

\begin{document}

\maketitle
\begin{abstract}

Estimating the health impacts of exposure to a mixture of chemicals poses many statistical challenges: multiple correlated exposure variables, moderate to high dimensionality, and possible nonlinear and interactive health effects of mixture components. Reviews of chemical mixture methods aim to help researchers select a statistical method suited to their goals and data, but examinations of empirical performance have emphasized novel methods purpose-built for analyzing complex chemical mixtures, or other more advanced methods, over more general methods which are widely used in many application domains. We conducted a broad experimental comparison, across simulated scenarios, of both more general methods (such as generalized linear models) and novel methods (such as Bayesian Kernel Machine Regression) designed to study chemical mixtures. We assessed methods based on their ability to control Type I error rate, maximize power, provide interpretable results, and make accurate predictions. We find that when there is moderate correlation between mixture components and the exposure-response function does not have complicated interactions, or when mixture components have opposite effects, general methods are preferred over novel ones. With highly interactive exposure-response functions or highly correlated exposures, novel methods provide important benefits. We provide a comprehensive summary of when different methods are most suitable.
\end{abstract}

\keywords{Chemical mixtures, epidemiology, simulation, hypothesis testing.}

\section{Introduction}

Observational studies analyzing the association between a mixture of multiple exposures and a health outcome face unique statistical challenges. Examples are \citeauthor{lebeaux2020maternal}\autocite{lebeaux2020maternal}, which studied the association between maternal serum concentrations of four per- and polyfluoroalkyl substances (PFAS) and thyroid hormone levels in maternal and cord serum, and \citeauthor{bobb2015bayesian}\autocite{bobb2015bayesian}, which used as a motivating example an analysis of the association between concentration of arsenic, manganese, and lead in umbilical cord blood and a psychomotor development score in early childhood. Including multiple relevant exposures in a model avoids left out variable bias, but the levels of different exposures may be correlated due to shared sources, making it difficult to distinguish effects. The number of exposures analyzed can be moderate to large, for example, 11 exposures in \citeauthor{carrico2015characterization}\autocite{carrico2015characterization}, to over 100 in \citeauthor{maitre2022state}\autocite{maitre2022state} Effects of exposures on health outcomes can be nonlinear and interactive, making model specification challenging.

A number of novel methods, designed specifically for analyzing chemical mixtures by dealing with one or more of these difficulties, have been proposed (see Section \ref{s:methods}). For our review, we define methods as ``novel" chemical mixture methods if they were designed specifically for this application, while we consider ``general" methods to be those in wide use in a variety of applications. Several reviews have summarized different methods, models, and considerations for researchers.\autocite{joubert2022powering, yu2022review} Several include simulation studies of properties of some methods, 
primarily novel chemical mixture methods, or other advanced methods which are not in wide use in applications.\autocite{hoskovec2021model, lazarevic2020performance, hao2025statistical} To our knowledge, recommendations so far have therefore either not included simulation studies, or included simulation studies but {did not include situations where simpler or more general methods may perform better than more complicated or novel methods}. Simulations so far have also not {treated Type I error rate control as a critical consideration}.\autocite{hoskovec2021model, lazarevic2020performance, hao2025statistical}
The resulting incomplete picture of mixture methods' empirical performance may bias researchers' choices of method. 

Our study assesses the performance of novel chemical mixture methods and more general methods, so that researchers may understand what method best meets their needs.
%
%
We compare methods' performance on Type I error rate control, power, and predictive accuracy, in an experimental manner across a range of simulated scenarios meant to model realistic situations.
We also examine qualitative aspects of model choice, such as additional assumptions and interpretability. { While we focus on exposure to chemical mixtures such as PFAS, our results are applicable to any study with similar statistical difficulties, such as studies involving multiple air pollutants or other correlated independent variables.}

Section \ref{s:setting} outlines the statistical setting; Section \ref{s:methods} describes several chemical mixture methods; Section \ref{s:sim} presents the simulation study design; Section \ref{s:results} presents results of the simulation; and Section \ref{s:discussion} presents our discussion and recommendations.

\subsection{Statistical setting}\label{s:setting}


The generic setting is trying to estimate parameters or characteristics of an exposure-response function $h$, where, letting $\mu_i \equiv E(y_i|\ba_i, \bx_i)$:
\begin{align}
    \eta_i=g(\mu_i) &= h(a_{ i1}, ..., a_{ ip}) + \mathbf{x}_i^T \boldsymbol{ \beta}_X,
\end{align}
where $a_{ij}$ denotes the value of exposure $j$ for observation $i$, $\mathbf{x}_i$ is the covariate vector for observation $i$, and $g$ is a monotonic, differentiable link function. With an assumed distribution for $y_i|\ba_i, \bx_i$, different choices of $g$ are used to model different types of data. Common cases are:
\begin{align}
    y_i &\sim N(\mu_i, \sigma^2), \quad \eta_i=g(\mu_i)=\mu_i, \label{e:gaussian}\\
    y_i &\sim Ber(\mu_i), \quad \eta_i = g(\mu_i) = \ln\left(\frac{\mu_i}{1-\mu_i}\right),\label{e:logistic} \\
    y_i &\sim Pois(\mu_i), \quad \eta_i=g(\mu_i) = \ln(\mu_i),\label{e:poisson}
\end{align}
where (\ref{e:gaussian}) models Gaussian-distributed continuous outcome data using the identity link function, (\ref{e:logistic}) models Bernoulli-distributed binary outcome data using the logistic link function, and (\ref{e:poisson}) models Poisson-distributed count outcome data using the log link function. For generality, we discuss models for the expected outcome on the scale of the link function, $\eta_i=g(\mu_i)$. 
All methods in Section \ref{s:methods} can model at least Gaussian and binary data.

We assume the exposure variables have continuous values, such as measured concentrations of chemicals in blood serum. For notational simplicity, we do not include covariates in the models.

\section{Statistical Methods for Chemical Mixtures}\label{s:methods}

Methods used for analyzing chemical mixtures may seek to accomplish any of three goals: 1) identifying elements of the mixture that are associated with the response, 2) identifying whether the mixture as a whole is associated with the response, and 3) {make predictions} on new (out-of-sample) data.

\subsection{General methods}
\subsubsection{Generalized linear models}\label{s:glm}

Generalized linear models (GLMs) assume that on the scale of the link function $g$, the expected value of the response is linearly related to the predictors. A GLM takes the form:
\begin{align*}
    \eta_i &= \beta_0 +  \mathbf{a}_i^T\boldsymbol{\beta}. 
\end{align*}
Nonlinearities and interactions must be modeled explicitly by the user, by including higher-order terms or transformations of the exposure variables as predictors in the model.
%
%
Correlation between exposures can be addressed using an F-test or $\chi^2$-test for joint association between any of the exposures and the response, testing $H_0\text{: } \beta_1 = \beta_2 = ... = \beta_p = 0$. {When covariates are included in the model, a partial F-test can be used to only test for association between the exposures and response, excluding covariate effects from the null hypothesis.}
Interpretation of GLMs is easy: $\beta_j$ represents the change in $\eta_i=g(\mu_i)$ per unit change in $a_{ij}$.
Functions to fit GLMs are included in R's base functionality.

\subsubsection{Principal Components Regression}

Principal Components Regression (PCR) reduces the dimension of the exposure vector $\mathbf{a}_i$ using principal components analysis (PCA) before regressing $y_i$ onto the new, derived explanatory variables.\autocite{hastie2009elements} In some cases, the new variables can represent underlying features of the data. In these cases, in high-dimensional problems, and where the assumption of linearity is reasonable, PCR may be useful. It also may be useful when the exposures are highly correlated because the derived variables are orthogonal. 
The original exposure variables are scaled, and then PCA obtains $M \leq p$ principal components (PCs) from the scaled exposures. The first PC explains the most variance in the scaled exposures, followed by the rest of the PCs in sequence. Often, a small number of PCs can approximate the scaled exposures.

Then the model  is
\begin{align}\label{e:pcr}
    \eta_i = \beta_0 + \mathbf{z}_{i}^T\boldsymbol{\theta},
\end{align} 
where the $M$-vector of PC values for observation $i$ is $\mathbf{z}_i$. Testing $H_0: \theta_1 = ...=\theta_M = 0$ tests for any association between the components of the mixture and the response. {When covariates are included, the dimension reduction step should not be applied to the covariates, and a partial F-test can be used to only test for association between the principle components and the response, excluding covariate effects from the null hypothesis.} This test does not account for uncertainty in determining the PCs or which PCs to include in the regression model. 

The PCs are weighted sums of the scaled exposure variables. If there are underlying features of the exposure data such as a shared source of multiple exposures, PCs may correspond to these features, although they must be interpreted by the user based on the exposure weights. It is not possible to conduct hypothesis tests for association of the individual chemicals with the outcome variable.

An exposure may be only lightly weighted in the selected PCs, but contribute substantially to the response, causing PCR to miss an important component of the exposure-response relationship, a risk not quantified by the fitted model. PCR is easily implemented using functions included in base R.

\subsubsection{LASSO/Elastic net regression}\label{s:enet}

LASSO regression and elastic net regression are linear models that perform variable selection rather than hypothesis testing, suiting them to high-dimensional settings when statistical inference is less important than prediction.\autocite{hastie2009elements} The model is again $\eta_i =\beta_0+ \mathbf{a}_i^T \boldsymbol{\beta}$. The LASSO estimator for identity link and Gaussian likelihood is

\begin{align}\label{e:lasso}
    \hat{\boldsymbol{\beta}}_{LASSO} = \arg \min_{\boldsymbol{\beta}} \left\{ \sum_{i=1}^n(Y_i -\beta_0 - \mathbf{a}_i^T\boldsymbol{\beta})^2 + \lambda \sum_{j=1}^p|\beta_{j}| \right\},
\end{align}
and the elastic net estimator is
\begin{align}\label{e:enet}
    \hat{\boldsymbol{\beta}}_{EN} = \arg \min_{\boldsymbol{\beta}}\left\{ \sum_{i=1}^n(Y_i -\beta_0- \mathbf{a}_i^T\boldsymbol{\beta})^2 + \lambda\left(\alpha\sum_{j=1}^p|\beta_{j}| + (1-\alpha) \sum_{j=1}^p\beta_{j}^2 \right)\right\},
\end{align}
with $\lambda>0, \alpha \in [0, 1]$. Typically cross-validation is used to select $\lambda$ and $\alpha$.\autocite{hastie2009elements} The penalty term $ \sum_{j=1}^p|\beta_{j}|$ in (\ref{e:lasso}) and (\ref{e:enet}) results in the $\beta_j$ being constrained to a $p$-dimensional region with corners along the coordinate axes; as a result the minima in (\ref{e:lasso}) and (\ref{e:enet}) may be attained at these corners, setting some of the $\beta_j$ to be zero and performing variable selection \autocite{hastie2009elements}.

Elastic net regression has sometimes been preferred in the analysis of exposure mixtures because LASSO has a tendency to arbitrarily select from among a group of correlated exposures, while elastic net tends to either exclude or include correlated exposure variables together \autocite{carrico2015characterization}. However, groupings under elastic net come from patterns of exposure and not associations with the response.
These methods' selections can be unstable \autocite{cadiou2021instability}. Coefficient estimates obtained after selecting variables are biased and typically do not reflect uncertainty in model selection \autocite{hastie2009elements, samanta2022estimation}. An R package used to fit elastic net models including LASSO, including for discrete outcome data, is \texttt{glmnet} \autocite{friedman2021package}.

\subsubsection{Generalized additive models}\label{s:gam}

Generalized additive models (GAMs) flexibly model the exposure-response relationship $h$ with simplifying assumptions. A GAM which is additive in the exposures is
\begin{align*}
    \eta_i = \beta_0 +\sum_{j=1}^pf_j(a_{ij}),
\end{align*}
where $\beta_0$ is the intercept, and $f_j, j= 1, ..., p$ is an unknown smooth function of the $j$th exposure. 
Typically the unknown functions are approximated using linear combinations of $k$ spline basis functions, so that the model is
\[f_j(a_{ij}) \approx \sum_{r=1}^kB_{jr}(a_{ij})\beta_{jr},\] with $B_{jr}(\cdot), r=1, ...,k$ denoting the B-spline basis functions used to approximate each function $f_j$ as in \citeauthor{wei2020sparse}\autocite{wei2020sparse}
Typically a smoothing penalty is applied when estimating the $\beta_{jr}$ to prefer smoother estimates of the $f_j$. GAMs can be interpreted by plotting the fitted functions. Including interactions is possible but often impractical except with low $p$.

Frequentist hypothesis tests for the statistical significance of smooth terms are available in software such as the R package \texttt{mgcv} \autocite{wood2011mgcv}. Per \citeauthor{wood2017generalized}, frequentist hypothesis tests may perform poorly when exposures are correlated, and better hypothesis test results are obtained by selecting the smoothing penalty using restricted maximum likelihood (REML).\autocite{wood2017generalized} Alternatively, penalties similar to those in LASSO (Section \ref{s:enet}) can be used to perform variable selection.
The implementations of GAMs in \texttt{mgcv} are highly optimized and computationally efficient.\autocite{wood2011mgcv}


\subsection{Novel methods for complex chemical mixtures}
\subsubsection{Bayesian kernel machine regression}

Bayesian kernel machine regression (BKMR) is a Bayesian implementation of Gaussian Process (GP) regression of the response onto the exposures.\autocite{bobb2015bayesian} The flexible GP model allows BKMR to estimate the exposure-response function even in the presence of nonlinear and interaction effects. BKMR uses a variable-selection prior to determine which components are associated with the response and estimates the model using Markov Chain Monte Carlo (MCMC). There are two approaches to variable selection available. The component-wise method performs variable selection on each individual mixture component. The hierarchical variable selection method performs variable selection on prespecified groups of mixture components, such as groups of highly correlated components. 

The Bayesian model for $\eta_i$ with component-wise variable selection is:
\begin{align}
    \begin{split}\label{e:bkmr}
        \eta_i &= h_i= h(a_{i1}, ..., a_{ip}),\\
        \mathbf{h} \equiv (h_1, ..., h_n)^T | \mathbf{K} &\sim N(\mathbf{0}, \tau\mathbf{K})\\
        \mathbf{K}_{\ell m} &\equiv K(\mathbf{a}_{\ell}, \mathbf{a}_m; \mathbf{r}) \equiv \exp \{-\sum_{j=1}^p r_j(a_{\ell j} - a_{m j})^2\}. \\
    \end{split}
\end{align}
The unknown exposure-response function $h$ is given a Gaussian Process prior with kernel matrix $\mathbf{K}$, defined through the kernel function $K(\cdot, \cdot)$ which determines the correlation between values of $h$ corresponding to different values of the exposures. 
BKMR scales the squared difference between between $a_{\ell j}, a_{mj}$ (the $\ell$th and $m$th observations of the $j$th exposure) by $r_j$ for each $\ell \neq m$, and variable selection is performed by giving $r_j$ a spike-and-slab prior with a spike at 0, for $j=1, ..., p$. 
When $r_j=0$, the $j$th exposure variable is deselected from the model, as it has no effect on $\mathbf{K}$ and therefore on the estimate of $h$. The posterior inclusion probability (PIP) for exposure $j$ is the posterior probability that $r_j$ is non-zero \autocite{bobb2015bayesian}. The hierarchical variable selection model yields a PIP for each group of variables, and then ranks the variables within each group by importance.

Much of BKMR's interpretation uses graphical summaries of the posterior distribution of $h$. Statistical inference is done most directly using the PIPs of the $p$ exposure components. Contrasts can also be estimated and tested using the posterior distribution of $h$, 
to test hypotheses related to joint or interaction effects of the different chemicals \autocite{bobb2018statistical}. Inference based on findings from graphical analysis of the posterior involves multiple comparisons, possibly many comparisons, and PIPs can be sensitive to prior specification. {BKMR is implemented in the R package \texttt{bkmr}.}

\subsubsection{Weighted quantile sum regression}

Weighted quantile sum (WQS) regression removes the effects of multicollinearity by considering the mixture as a whole.\autocite{carrico2015characterization} WQS regression forms a weighted sum, or index, of the exposure variables, and regresses the outcome onto the index. WQS regression uses the quantized versions of the exposure variables as inputs to the index, termed the ``weighted quantile sum." Let $q_{ij}$ denote the values of the quantized exposure variables. For example, $q_{ij}$ would be the quantized value of $a_{ij}$, so that if sample quartiles are used, $q_{ij}$ could take values 1, 2, 3, or 4. The WQS model is
\begin{align}
    \eta_i = \beta_0 + \beta_1 \bigl(\sum_{j=1}^pw_jq_{ij}\bigr) ,
\end{align}
where $w_j$ is the weight assigned to exposure variable $j$. Inference on the association of the mixture with the response is performed by testing $H_0$: $\beta_1=0$. The sample is by default split into training and validation sets, with the training set used to estimate the weights, and the validation set used to test $H_0$. 
WQS regression assumes that all exposures have either non-decreasing or non-increasing associations with the outcome, an assumption sometimes termed ``directional homogeneity." \autocite{keil2020quantile} 
WQS regression is implemented in the R package \texttt{gWQS}.


\subsubsection{Quantile g-computation}

Similar to WQS regression, quantile g-computation (QGC) assesses the association of the entire mixture with the response.\autocite{keil2020quantile} The model is
\begin{align}\label{e:qgc}
    \eta_i &= \beta_0 + \sum_{j=1}^p\beta_{j}q_{ij} ,
\end{align}
where the notation is as above. Inference is performed on the quantity $\psi = \sum_{j=1}^p\beta_j$, interpreted as the expected increase in the outcome when increasing the level of every mixture component by one quantile. Model (\ref{e:qgc}) can be extended to include nonlinear or product terms of the exposures. Weights for each exposure are defined by separately normalizing the positive and negative regression coefficient estimates to sum to 1, which can be misleading.
%
QGC regression is implemented in the R package \texttt{qgcomp}.

\section{Simulation Study}\label{s:sim}
In our simulation study, different methods are evaluated on: 1) their ability to control Type I error probability, 2) their power to detect association of an individual component with the response, 3) their power to detect association of the mixture with the response, and 4) predictive accuracy measured by mean squared error (MSE) on new data. We evaluated four general methods and three novel mixture methods, including several variations, on these criteria.
Details of the methods' implementation are in the Supplementary Materials section S1. Abbreviations used in results tables and figures are listed in Table \ref{tab:abbrs}.

\begin{table}[]
    \centering
    \begin{tabular}{p{.2\linewidth}p{.75\linewidth}}
         Abbreviation &  Method and notes \\
         \hline
         GLM & Generalized linear model, using individual hypothesis tests to test significance of individual components or F-test to test significance of the mixture. Used with identity link function and Gaussian outcome distribution, i.e., ordinary least-squares regression. \\
         ENET & Elastic net, using variable selection to select individual components. \\
         PCR & Principal components regression, using an F-test to test the significance of the mixture, selecting enough components to explain {75}\% of the overall variance in exposures.\\
         GAM (HT) &  GAM using frequentist hypothesis testing to test significance of individual components. \\
         GAM (VS) & GAM using variable selection to select individual components. \\
         GAM (HT, Bonf.) & GAM using Bonferroni's correction of individual component hypothesis tests to test significance of the mixture. \\
         \hline 
         BKMR (Contrast) & BKMR testing for the significance of the mixture using an estimated contrast between levels of overall exposure. \\
         BKMR (Hier, 0.95) & BKMR testing for the significance of the mixture using hierarchical variable selection, and a PIP at least 0.95 indicating significance. \\
         BKMR (0.50) & BKMR testing the significance of individual components, and a PIP at least 0.50 indicating significance. \\
         BKMR (0.95) & BKMR testing the significance of individual components, and a PIP at least 0.95 indicating significance.\\
         QGC & Quantile g-computation, testing the significance of the mixture. \\
         WQS &  Weighted quantile sum regression, testing the significance of the mixture, assuming positive associations with all individual components.
    \end{tabular}
    \caption{Abbreviations for methods used in the simulation study. General methods are listed in the top half of the table, followed by novel chemical mixture methods.}
    \label{tab:abbrs}
\end{table}


\subsection{Data generating process}\label{s:data_gen}

The sample size $n$ was either $100$ or $400$, and the number of exposure variables $p$ was either 5, 10, or 20. The exposures were generated randomly such that each had an Exp(1) marginal distribution, and the correlation $\rho$ between each pair of exposures was 0, 0.5, or 0.9.


The response variable $y_i$ was generated as:
\begin{align*}
    y_i &= h(\ba_i; \beta) + \epsilon_i \\
    \epsilon_i &\sim N(0, 2^2), \text{i.i.d.}
\end{align*}
In the main simulations, the true exposure-response function $h$ was linear, nonlinear, linear-interactive, or sinusoidal (which is both nonlinear and interactive):
\begin{align*}
    h_{lin}(\ba; \beta) &= \beta a_1 + \beta a_2 \\
    h_{nonlin}(\ba; \beta) &= \beta a_1^2 + \beta a_2^2 \\
    h_{int}(\ba; \beta) &= \beta a_1a_2  \\
    h_{sine}(\ba; \beta) &= \beta \sin\left( \frac{\sqrt{a_1^2 + a_2^2}}{2} \right).
\end{align*}
In these main scenarios, only two exposure variables were related to the response regardless of the value of $p$. In an additional scenario we used $h_{opp}(a_1, a_2; \beta) = \beta a_1 +  a_2$ and varied $\beta$ from -1 to 1, so that components' effects could cancel out. In a final ``dense" scenario, all exposures affected the response, with $h_{dense}(a_1, ..., a_p; \beta) = \mathbf{a}^T\boldsymbol{\beta}$ where $\boldsymbol{\beta} = (\beta, ..., \beta)^T$. We tested two hypotheses to evaluate different methods, $H_0$: $a_1$ unassociated with $y$, and $H_0$: no association between the mixture and response. 

Values of $\beta$ ranged from 0 to an endpoint informative for each scenario. Prediction mean-squared error was assessed using $100$ new points drawn from the same distribution as the samples used for model fitting, except without the noise term $\epsilon_i$ added. For $n=100$, we used 400 Monte Carlo (MC) iterations for each value of $\beta$ in each scenario, and for $n=400$, we used 200 MC iterations. In each scenario, the estimated Type I error rate was the proportion of MC iterations in which the method wrongly determined the component or mixture was associated with the response when $\beta=0$. { For frequentist hypothesis tests, if p-values were less than the nominal value $\alpha=0.05$, the null hypothesis was rejected; decision rules for other methods are described in Table \ref{tab:abbrs}.} Nominal Type I error rate for methods based on frequentist hypothesis testing was { therefore} $\alpha=0.05$. The estimated power for each value of $\beta\neq0$ was the proportion of MC iterations in which the method correctly determined that the component or mixture is associated with the response.  

\section{Results}\label{s:results}

Results showed generally that when correlation between exposures, number of exposures, and/or complexity of the exposure-response function were high, then the novel methods had higher power than general methods. In scenarios where correlation was moderate or low, there were fewer exposures, and the exposure-response function was not too complicated, the general methods were more powerful. Method recommendations based on different scenarios or priorities for the data analysis are presented at the end of this section in Table \ref{tab:summary}. Detailed summary tables of the strengths and weaknesses we found of each method are in Tables {\ref{tab:summary2a} and \ref{tab:summary2b}}.

{A selection of simulation study results are included in Figures \ref{fig:power_varsel_n100_p5_rho5} - \ref{fig:mse_n100_p5_rho5}.} Remaining results are in the Supplemental Materials Figures S1-S27 and Tables S1-S16. The power curves display the probability of rejecting the null hypothesis on the y-axis, for the value of $\beta$ on the x-axis. When $\beta=0$, the value of the curve is the Type I error probability (targeting 0.05) since when $\beta=0$ there is no association between the exposures and response. Therefore, the power curve for a method that outperforms competitors has a y-intercept at maximum 0.05, and then the rest of its curve above those of other methods. The prediction error curves display the prediction mean squared error (MSE), using predictions on new data not included in the original sample, for the value of $\beta$ on the x-axis. 

The power curves for component-wise hypothesis tests in Figure \ref{fig:power_varsel_n100_p5_rho5} show that elastic net and BKMR, using a PIP cutoff of 0.50 to determine significance, and {GAMs using variable selection, abbreviated as GAM (VS),} were over-sensitive, resulting in Type I error rates well above 0.05. {GAMs using hypothesis testing, abbreviated as GAM (HT),} GLMs, and BKMR using a PIP cutoff of 0.95 had Type I error rate at or below the nominal value of 0.05. {GAM (HT)} and GLMs had similar performance in the linear, nonlinear, and interactive scenarios, with greater power than BKMR using a PIP cutoff of 0.95. In the sinusoidal scenario, however, BKMR using a PIP cutoff of 0.95 strongly outperformed {GAM (HT)} and GLMs in power, reflecting the more flexible model's ability to estimate the complicated exposure-response function. The prediction MSE curves in Figure \ref{fig:mse_n100_p5_rho5} show generally that more accurately-specified models performed better on prediction MSE when the association between mixture and response is strong.

\begin{figure}
    \centering
    \includegraphics[width=0.9\linewidth]{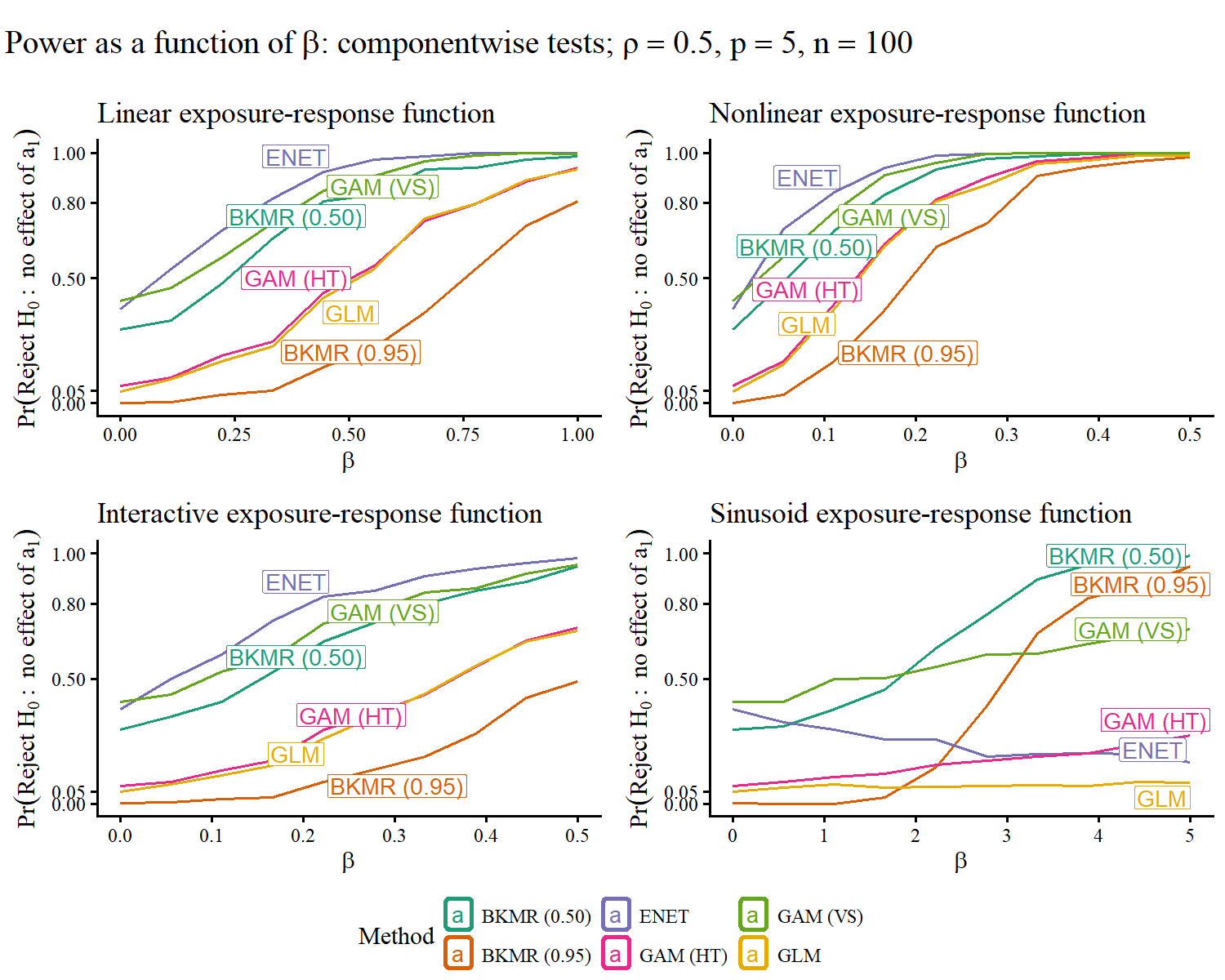}
    \caption{For each method (curve), the y-value is the probability of rejecting $H_0$: first mixture component $a_1$ not associated with the response, for different strengths of association $\beta$ (x-axis). When $\beta=0$, the value of the curve is the estimated Type I error probability. The targeted Type I error rate was $0.05$. When $\beta \neq 0$, the value of the curve is the power of the hypothesis test. Better methods have Type I error rate at most 0.05, and power curve above other methods. Exposure-response functions used are linear, nonlinear, linear interaction, and sinusoid (nonlinear interaction).}
    \label{fig:power_varsel_n100_p5_rho5}
\end{figure}

\begin{figure}[H]
    \centering
    \includegraphics[width=0.9\linewidth]{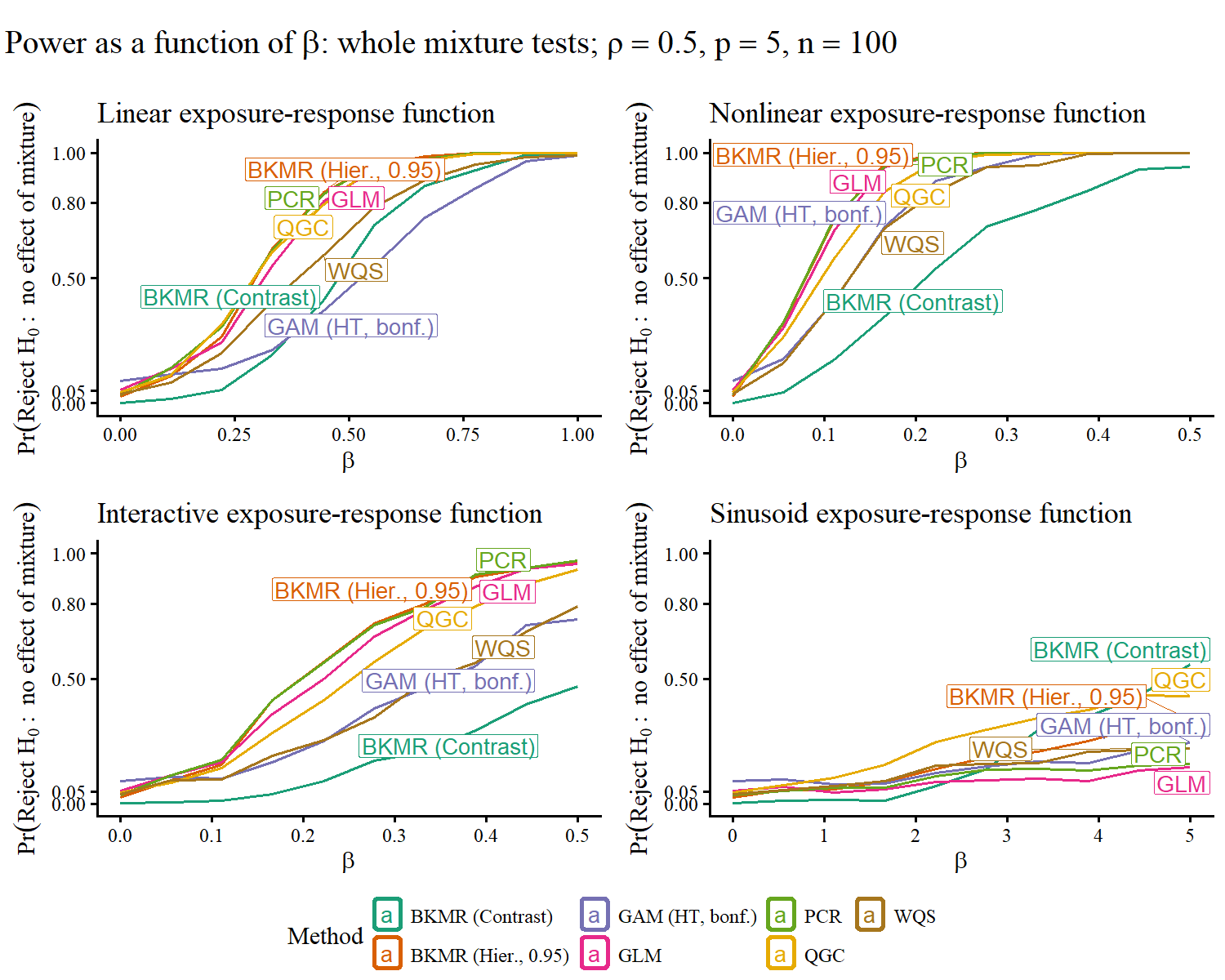}
    \caption{For each method (curve), the y-value is the probability of rejecting $H_0$: the overall mixture is not associated with the response, for different strengths of association $\beta$ (x-axis). When $\beta=0$, the value of the curve is the estimated Type I error probability. The targeted Type I error rate was $0.05$. When $\beta \neq 0$, the value of the curve is the power of the hypothesis test. Better methods have Type I error rate at most 0.05, and power curve above other methods. Exposure-response functions used are linear, nonlinear, linear interaction, and sinusoid (nonlinear interaction).}
    \label{fig:power_idx_n100_p5_rho5}
\end{figure}

\begin{figure}[H]
    \centering
    \includegraphics[width=0.9\linewidth]{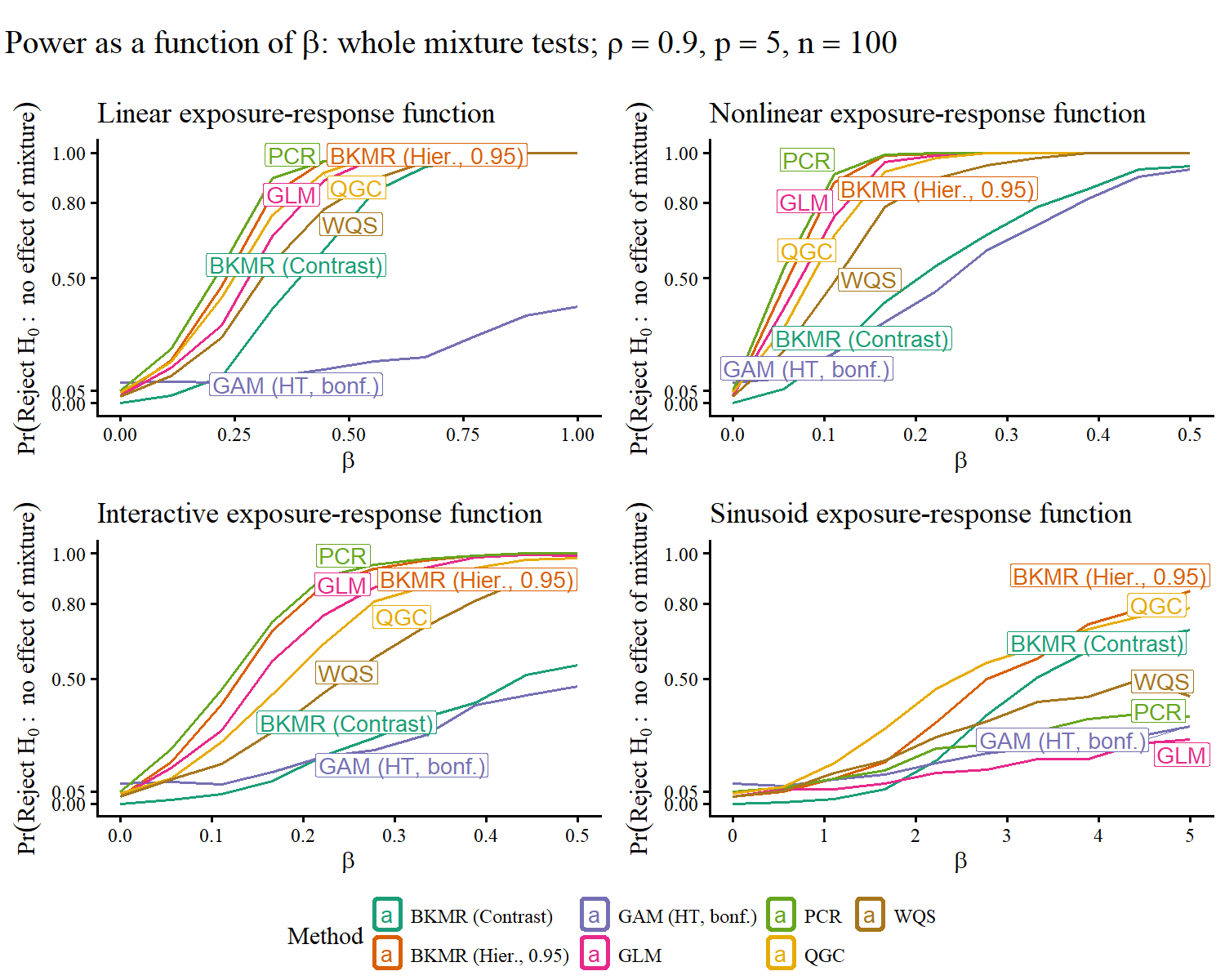}
    \caption{For each method (curve), the y-value is the probability of rejecting $H_0$: the overall mixture is not associated with the response, for different strengths of association $\beta$ (x-axis). When $\beta=0$, the value of the curve is the estimated Type I error probability. The targeted Type I error rate was $0.05$. When $\beta \neq 0$, the value of the curve is the power of the hypothesis test. Better methods have Type I error rate at most 0.05, and power curve above other methods. Exposure-response functions used are linear, nonlinear, linear interaction, and sinusoid (nonlinear interaction).}
    \label{fig:power_idx_n100_p5_rho9}
\end{figure}


\begin{figure}[H]
    \centering
    \includegraphics[width=0.5\linewidth]{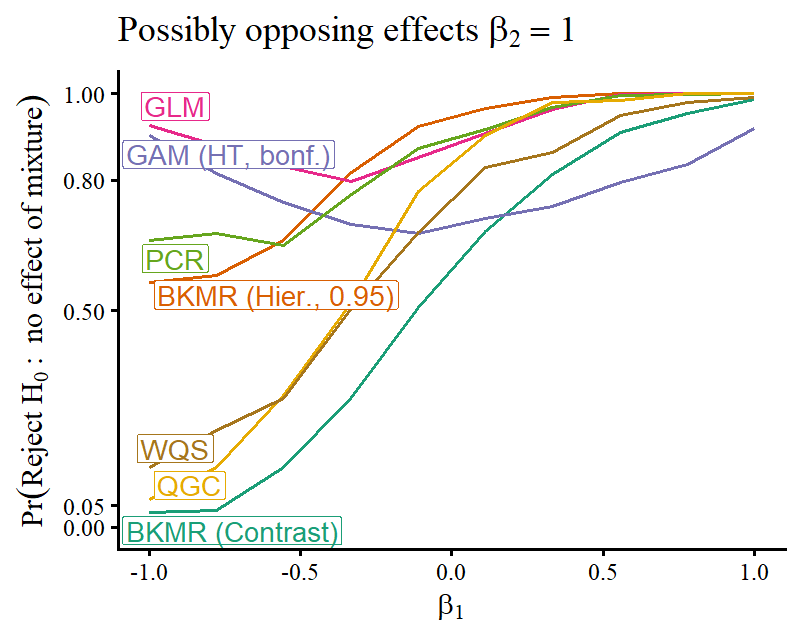}
    \caption{Power curves for whole-mixture hypothesis tests for $n=100, p=10, \rho=0.5$, possibly opposing effects. For each method (curve), the y-value is the probability of rejecting $H_0$: the overall mixture is not associated with the response, for different strengths of association $\beta$ (x-axis). Exposure-response function used is linear with possibly opposing effects of mixture components. Note that the x-axis endpoints have changed as $E(y_i|a_{1i}, a_{2i}) = \beta_1a_1 + \beta_2 a_2$, and $\beta_2$ is fixed at 1, and $\beta_1$ now varies from $-1$ to $1$. Therefore better methods have all values of their power curve above other methods'.}
    \label{fig:power_idx_n100_p10_rho5_opp}
\end{figure}

\begin{figure}
    \centering
    \includegraphics[width=0.9\linewidth]{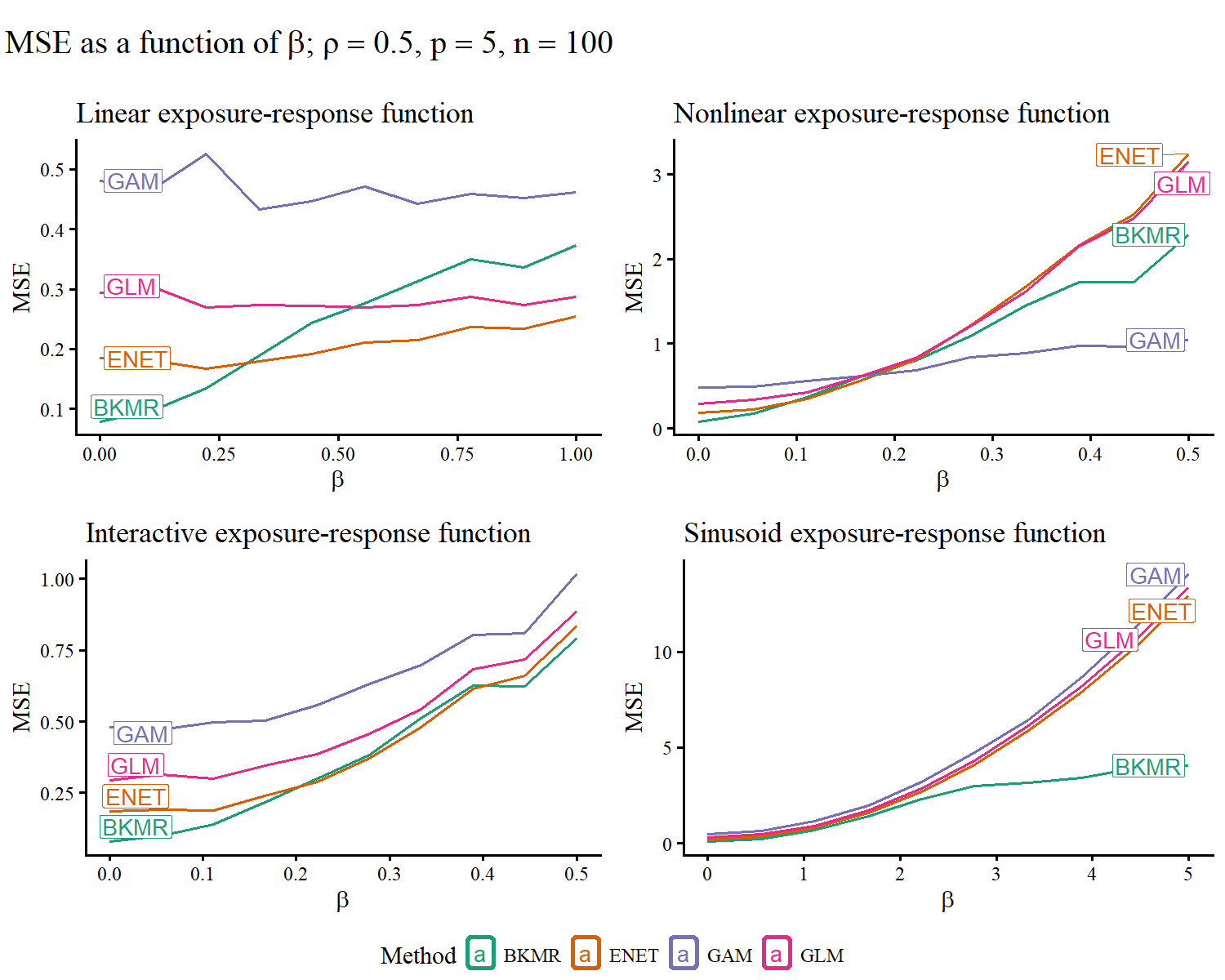}
    \caption{For each method (curve), the y-value is the prediction mean squared error (MSE) on new data, for different strengths of association $\beta$ (x-axis). Better methods have lower MSE curves. The new data are generated from the same exposure-response functions as the data used to fit the model, but have no random variation added. Exposure-response functions used are linear, nonlinear, linear interaction, and sinusoid (nonlinear interaction).}
    \label{fig:mse_n100_p5_rho5}
\end{figure}


\begin{table}[]
    \centering
    \begin{tabular}{ll}
         Analysis priority & Preferred method(s) \\
         \hline
         High correlation between exposures and/or many exposures & QGC, WQS, PCR \\
         Moderate correlation between exposures and/or moderate number of exposures & GLMs, GAMs \\
         Nonlinear exposure-response function & GAMs\\
         Interactive exposure-response function & BKMR \\
         Interpretable results & GLMs, QGC \\
         Opposite effects of exposures & GLMs, GAMs \\
         Low sample size & All but BKMR
    \end{tabular}
    \caption{Preferred methods for various analysis priorities that may arise in analysis of chemical mixtures' associations with health outcomes. Analysis priorities may be dealing with particular challenges posed by the data, or more qualitative priorities such as the need for interpretable results.}
    \label{tab:summary}
\end{table}

\subsection{Mixture component identification}

GAMs and GLMs worked best in {many} situations for identifying important components of the mixture, with nominal Type I error rate and competitive power. This was true when misspecified due to nonlinearity (in the case of GLMs) or linear interaction, but was not true for the most complicated, sinusoidal, exposure-response relationship (Figure \ref{fig:power_varsel_n100_p5_rho5}). 
%
When the form of the interaction was most complicated in the sinusoidal scenario, 
only BKMR was easily able to detect the association.

The favorable results for GLMs (specified as linear models) are in part due to the simulation set up. The nonlinear relationships were monotonically increasing, so a misspecified linear model will fit an increasing trend to the data. Our nonlinear and linear interaction scenarios which were chosen to represent exposure-response functions that were more complicated than linear models but still relatively simple.

Elastic net, GAMs with variable selection, and BKMR with a PIP cutoff of 0.5 had high Type I error rates that would generally be unacceptable for hypothesis testing, but may be useful for more exploratory analyses. BKMR using the higher cutoff of 0.95 had acceptable Type I error rate, although lower power than other methods in simpler scenarios.

\subsection{Whole-mixture tests}
The GLM F-test is suitable for moderate $p$ {and $\rho$}, but loses power compared to methods that perform whole-mixture tests that use fewer degrees of freedom, such as QGC, WQS, and PCR, {with higher $p$ or $\rho$. Figures \ref{fig:power_idx_n100_p5_rho5} and \ref{fig:power_idx_n100_p5_rho9}, top-left panels, illustrate that QGC and PCR outperform GLMs with increased $\rho$ in the linear scenario; the effect of increasing $p$ and/or $\rho$ can be further seen in Figures S9 - S14 in the Supplemental Materials, and we discuss how to interpret these results in the context of applied data analyses in Section \ref{s:discussion}}. When associations went in opposite directions, QGC, WQS, PCR, and BKMR's whole-mixture tests lost power{, with the loss for QGC, WQS, and BKMR's contrast test particularly severe} ({Figure \ref{fig:power_idx_n100_p10_rho5_opp}}). Power for WQS regression was generally lower than QGC, due at least in part to splitting used by WQS. In the main simulation scenarios {where $p=5$}, GAMs using the Bonferroni correction worked well { when $\rho=0$ (Figure S9), had acceptable power when $\rho=0.5$ (Figure \ref{fig:power_idx_n100_p5_rho5}), but had uncompetitive power when $\rho=0.9$ (Figure \ref{fig:power_idx_n100_p5_rho9}); however, in the opposite-effects scenario, power was still greater than methods other than GLMs (Figure \ref{fig:power_idx_n100_p10_rho5_opp}). When $p=10$, increasing $\rho$ decreased GAMs' power more dramatically when the Bonferroni correction was applied (Figures S12-S14).} 

BKMR's default hierarchical variable selection prior resulted in group PIPs that were essentially always greater than 0.50 even under the null hypothesis, displayed in Figure S29 in the Supplemental Materials. 
However, when the cutoff for significance was 0.95, empirical performance for hierarchical variable selection was strong in most scenarios, with some power lost in the opposite effects scenario. The BKMR contrast test had appropriate Type I error rate control, but low power compared to parametric methods in the simpler scenarios.


\subsection{Prediction}

Prediction mean squared error favored the correctly-specified model when strength of association between the exposures and response, controlled by $\beta$, was higher. In the linear scenarios, GLM and elastic net were most favored; in the nonlinear scenarios, GAMs were favored followed closely by BKMR; and in the linear interaction and sinusoidal scenarios, BKMR was consistently best (Figure \ref{fig:mse_n100_p5_rho5} is representative, with full results in Figures S18-S27). When $\beta$ was low, elastic net and BKMR were best even in some scenarios where other models were either better-specified or more simply specified. This is likely due to their variable selection procedures removing irrelevant variables from the model, which seems most important for low $\beta$.
\begin{table}
    \centering
    \small
    \begin{tabular}{p{0.12\linewidth}|p{0.2\linewidth}|p{0.2\linewidth}|p{0.2\linewidth}|p{0.2\linewidth}|}
         Method&  Nonlinearity, interactivity&  Multiple correlated exposures&  Identify individual components&  Overall Performance\\
         \hline
         \hline
         Generalized linear model (GLM)&  Poor: user-specified model.&  OK: joint tests; good with low-moderate $p$ and $\rho$.&  Good if $p$ not too high.&  Good if model approximately correct.\\
         \hline
         Generalized additive model (GAM)&  Good for nonlinearity; interactivity impractical except for low $p$.&  Poor: loss of power with correlated predictors. &  Good if $p$ not too high.&  Good.\\
         \hline
         Elastic net (ENET) (including LASSO)&  Poor: user-specified model. &  OK: variable selection good for sparse effects, but ENET tends to (de)select correlated groups, LASSO will select within groups arbitrarily; does not control Type I error.&  Excellent: Unimportant components deselected, but no uncertainty quantification.&  Poor Type I error control. If that is unimportant, good performance.\\
         \hline
         Weighted quantile sum (WQS) &  OK: quantized variables may implicitly model nonlinearity; user-specified model.&  Excellent if only interested in whole mixture; requires directional homogeneity.&  OK: Gives weights on each component, but no statistical testing. &  Good if assumptions met, but loses power by splitting sample.\\
         \hline
         Quantile g-computation (QGC) &  OK: quantized variables may implicitly model nonlinearity; user-specified model.&  Excellent if only interested in whole mixture; suffers without directional homogeneity.&  Good if using individual regression coefficients. Outputted weights may be misleading.&  Good if assumptions met; does not split the sample.\\
         \hline
         Bayesian kernel machine regression (BKMR) &  Excellent: if enough signal/data.&  OK: use hierarchical model PIPs with caution. &  OK: PIPs are sensitive to prior, appropriate cutoff for significance is unclear, should be used with care.&  Good with interaction effects. PIP cutoff of 0.50 oversensitive. Default prior for hierarchical selection is poorly calibrated. \\
         \hline
         Principle components regression (PCR)&  Poor: user-specified model.&  Good, as it combines exposures; however, the combination ignores the outcome, so can ignore important components.&  Poor: the principle components are difficult to interpret except in limited circumstances.&  Good for whole-mixture tests, although will fail to identify components not included in the PCs used.\\
         \hline
    \end{tabular}
    \caption{Detailed summary of different mixture methods' empirical performance. $p$ denotes the number of exposure variables, and $\rho$ denotes their pairwise correlations.}
    \label{tab:summary2a}
\end{table}

\begin{table}
    \centering
    \small
    \begin{tabular}{p{0.12\linewidth}|p{0.2\linewidth}|p{0.25\linewidth}|p{0.15\linewidth}|p{0.2\linewidth}|}
         Method&  Interpretability& Limitations & R Implementation & When to use\\
         \hline
         \hline
         Generalized linear model (GLM)&  Excellent: coefficients have clear meaning if assumptions satisfied.&  Strict parametric assumptions.& Base R & Low-to-moderate $p$, so specifying a model is reasonable.\\
         \hline
         Generalized additive model (GAM)&  Good: can easily examine plots of fitted functions.&  No joint test, limited ability to model interaction.& mgcv & Nonlinearity present, but whole-mixture test unimportant.\\
         \hline
         Elastic net (ENET) (including LASSO)&  OK: coefficients have meaning, but are biased and lack uncertainty quantification.&  Poor Type I error control; no uncertainty quantification; user-specified model.&glmnet; caret for hyperparameter tuning. & Many predictors, and Type I error control unimportant.\\
         \hline
         Weighted quantile sum (WQS) &  OK: Index effect difficult to interpret, but weights have clear meaning.&  User-specified model; directional homogeneity requirement; limited interpretability; sample splitting.&gWQS & Moderate-to-many correlated predictors and individual components unimportant, directional homogeneity.\\
         \hline
         Quantile g-computation (QGC) &  OK: Index effect interpretable, but weights misleading; users can interpret initial regression fit.&  Suffers without directional homogeneity; user-specified model.&qgcomp & Moderate-to-many correlated predictors and individual components unimportant, directional homogeneity.\\
         \hline
         Bayesian kernel machine regression (BKMR) &  Good qualitatively: detailed graphical summaries of posterior. OK quantitatively: can estimate contrasts that may be crude summaries.&  Nonparametric: prefers larger $n$; PIPs sensitive to prior tuning and lack guidance on cutoff; examining posterior involves multiple comparisons.&bkmr& Lot of data or strong signal, or improved (cross-validated) predictions over other methods.\\
         \hline
         Principle components regression (PCR)&  OK: In some situations, PCs may correspond to features of exposure data. Otherwise, poor.&  Dimension reduction ignores relationship with outcome; poor interpretability.&Functions required are in base R. & Many components; individual components unimportant or a PC is meaningful; can risk missing an important components.\\
         \hline
    \end{tabular}
    \caption{Detailed summary of different mixture methods' interpretability, limitations, software, and recommended use. $p$ denotes the number of exposure variables (mixture components), and $\rho$ denotes their pairwise correlations.}
    \label{tab:summary2b}
\end{table}

\section{Discussion}\label{s:discussion}


In our simulated scenarios where the number of exposures was moderate (in our simulations, $p=5$), the correlation between exposures was moderate (in our simulations, $\rho=0.5$), and the exposure-response function was relatively simple (i.e., not the sinusoid function in our simulations), general methods outperformed novel chemical mixture methods in hypothesis testing. Novel chemical mixture methods should therefore not be a default choice. The choice of model and method should be guided by the researcher's knowledge of the data and their goals, but the tradeoffs of methods in different scenarios may not have been well-understood to this point due to the lack of simulation studies comparing general methods and novel chemical mixture methods across a range of scenarios. Unless there is something particularly difficult about analyzing a particular data set involving chemical mixture exposures, general methods such as GLMs and GAMs will likely be better choices for the analysis than a more specialized method.

{

Previous simulation studies have not evaluated performance of simpler methods applied to simpler (e.g., additive and linear) data generating processes. \citeauthor{hao2025statistical} considers 11 methods, comprised of variants of elastic net/LASSO regression, BKMR, random forests,\autocite{breiman2001random} as well as QGC, WQS regression, selection of nonlinear interactions using forward selection (SNIF),\autocite{narisetty2019selection} and an ensemble method.\autocite{vanderLaanPolleyHubbard+2007} \citeauthor{hao2025statistical} recommended SNIF for identifying important mixture components when Type I error rate control was of primary interest.
\autocite{hao2025statistical} \citeauthor{lazarevic2020performance} considered six nonlinear regression methods that perform variable selection, and compared them primarily using the F1-statistic, which balances Type I and II error rates. \citeauthor{lazarevic2020performance} found that several nonlinear regression methods were approximately equally suitable for identifying important mixture components, including BKMR using a PIP cutoff of 0.50 to determine importance of mixture components.\autocite{lazarevic2020performance} \citeauthor{hoskovec2021model} compared an interactive Bayesian linear regression model with nonparametric shrinkage,\autocite{herring2010nonparametric} a clustering model,\autocite{molitor2010bayesian} BKMR, and linear regression using frequentist hypothesis tests, but examined only interactive data-generating processes. \citeauthor{hoskovec2021model} recommended the interactive Bayesian linear model using the shrinkage prior when identifying important components was of primary interest and the truth was approximately linear.\autocite{herring2010nonparametric, hoskovec2021model}
%
%
In contrast to these previous studies, we clearly identify scenarios where simpler methods like additive GLMs (e.g., linear regression), or GAMs with an additive specification, provide more power than more complicated or novel methods while controlling the Type I error rate to a chosen, nominal level.
}

Our results show that novel methods for chemical mixtures have advantages in hypothesis testing performance when there is high correlation (in our simulations, $\rho=0.9$) between exposures or many (in our simulations, $p=10$) moderately-correlated exposures (QGC and WQS); or the dose-response function is very nonlinear and interactive (BKMR). There is no clear dividing line between scenarios that allows a definitive, {\it a priori} best choice of method for a particular data set. Considering linear (on the scale of the link function $g$) exposure-response relationships, in determining what method might be best, there is interplay between the number of exposure variables, their correlations, the strength of association between the exposures and the response, and the sample size. More variables and higher correlations will tend to reduce the performance of GLMs, while a {higher} sample {size} and/or strength of association will increase the performance of GLMs. For a data set with many highly correlated exposure variables and a smaller sample size, QGC or WQS are appropriate as GLMs may not be able to discern individual associations of the correlated exposures with the response, but if a much larger sample size is obtained, GLMs may then perform better, negating the need for a more specialized method. It can also be difficult to determine whether the true exposure-response function is complicated enough to warrant using BKMR, especially if the overall strength of association is not high. Predictions can be obtained on the data set using different models, using cross-validation to avoid overfitting, in order to compare the predictive accuracy of each model for those data.\autocite{hastie2009elements} If BKMR or another flexible method provides notably better predictive accuracy than a simpler model, then the more flexible model may make more sense; standard GLM diagnostic plots may suggest model changes that achieve the same goal, however.


Our simulation design was meant to represent aspects of real data sets researchers may find themselves analyzing. For example, although the sample sizes of our simulations were limited to $100$ and $400$, and the number of exposures was $5, 10$, or $20$, results from these combinations should represent, roughly, results from other data sets with similar ratios of number of exposures to sample size. Our results for $p=5$ and $n=100$, for example, we believe are representative of results that could be expected of other data sets where $p/n \approx 0.05$. Similarly, although we considered the strength of association controlled by the regression parameter $\beta$ in simulations, with variance of the i.i.d. error terms fixed at 4, each result is likely representative of other scenarios where the signal-to-noise ratio $Var(h(\ba; \beta))/Var(\epsilon_i)$ is preserved with other choices of parameters. However, the simulations do not capture a problematic scenario for PCR, when important components of the mixture are not included in selected PCs. There are certainly many other real-life scenarios not represented in our simulation. {For example, with extremely large $n$, most methods will have adequate power to detect practically significant exposure-response associations, and the primary factor in choosing a method may then be interpretability with regard to specific research questions.}

There are important tradeoffs to using each of the methods. In general, to address correlation between exposures, their associations with the response are combined in some way in the novel methods and PCR. Although only WQS makes the explicit ``directional homogeneity" assumption, WQS, QGC, PCR, and BKMR's whole-mixture tests { lost power} when effects went in opposite directions in our simulations. Without directional homogeneity, then GLMs or GAMs are likely preferable.

Both WQS and QGC require quantization of continuous exposure variables. This is also sometimes performed with other methods such as GLMs. This can help model nonlinear relationships between the exposures and response and reduce influence of outliers.
Quantization loses information from the exposure data\autocite{yu2022review}. Near the values of sample quantiles, it can create large jumps in values in the independent variable $q_{ij}$, when in reality there was a small difference between the original $a_{ij}$ values, or within one quantile, very different values of $a_{ij}$ may be given the same value of $q_{ij}$. The locations of the changes in value are essentially arbitrary. Quantizing continuous exposure variables should therefore be done with caution, and when possible{, other methods to model nonlinearity should be preferred.}

In \citeauthor{Gennings_2021}, QGC was criticized for misunderstanding one goal of WQS: to estimate effects of many small exposures together.\autocite{Gennings_2021} In our ``dense" scenarios, capturing this situation, WQS had modestly worse power than QGC and several other methods (Table S17, Figure S18). This was likely due in part to the sample-splitting used by WQS. If many small effects exist, if they are all in the same direction, our results showed QGC to be the most appropriate choice. \citeauthor{yu2022review} also notes that WQS regression lacks interpretability since $\beta_1$ does not have an easily-expressed meaning.\autocite{yu2022review} In both WQS and QGC the index is a linear combination of the exposures.

BKMR's PIPs require careful interpretation. Using 0.50 as the cutoff for statistical significance led to high Type I error rates in our simulation. With hierarchical variable selection, group PIPs were nearly always above 0.50 even under the null hypothesis of no association of the mixture with the response, when using default {package} settings (Figure S29). Using a cutoff of 0.95 resulted in Type I error rates less than 0.05 in our simulations, although this cutoff was chosen essentially arbitrarily and may not be appropriate for all situations. PIPs can also be influenced by the prior distributions, and sensitivity to the priors should be checked as recommended in \citeauthor{bobb2015bayesian}\autocite{bobb2015bayesian} For these reasons, it is difficult to interpret the strength of evidence of association indicated by a PIP unless it is close to 1.0. 

As indicated by the high PIPs under the null hypothesis when using hierarchical variable selection ({Figure S29}), prior inclusion probabilities need to be tuned according to the number of chemicals in the group. While this tuning requirement is a reasonable component of such a grouped variable selection procedure, it is not mentioned in \citeauthor{bobb2015bayesian} or \citeauthor{bobb2018statistical}, and with inappropriate package default settings, raises the possibility of accidental misuse by practitioners.\autocite{bobb2015bayesian, bobb2018statistical} Predictions can be used to explore a fitted BKMR model and draw conclusions from the analysis. Conclusions based on exploration of the model predictions' posterior distribution should be differentiated from those based on statistical tests that have appropriate control of Type I error rates. 

The traditional methods also require tradeoffs. GLMs require a user-specified dose-response function (although this is also true of QGC and WQS). PCR can miss important components of the mixture in the dimension reduction step and also assumes a linear model. GAMs lack a good joint test of association, making them unsuitable when there is too much correlation between exposures. In general, GLMs and GAMs should be used when there are not a large number of exposure variables relative to the sample size, and the correlation between them is moderate.

Finally, many of the methods considered, including general ones, often sacrifice some interpretability, which is important when communicating results to community members or policymakers. GLMs provide excellent interpretability, although there are scenarios in which they are unsuitable. {Individual component effects can be read directly from the GLM model output, and mixture effects and confidence intervals for GLMs or GAMs could be estimated using contrasts between different exposure values, such as contrasting predicted values when each exposure is set to its 0.75 quantile versus its 0.25 quantile.} One useful strategy may be to use GLMs to obtain interpretable summaries of results, and check robustness of the overall results to the strong GLM assumptions using a more complicated model such as GAMs, or a novel mixture method, as was done in \citeauthor{ROSATO2024120082} and \citeauthor{lebeaux2020maternal}, for example.\autocite{ROSATO2024120082, lebeaux2020maternal}

\section{Conclusion}

As interest in modeling chemical mixtures continues to grow, more innovative techniques for analyzing these types of data will be proposed. The challenges in this field are significant and we hope will continue to spur new statistical developments. Novel approaches hold great promise to further our ability to understand health effects of environmental contaminants or other types of chemical mixtures, and will advance other areas of statistical methodology. Given enthusiasm for new approaches, it is critical to evaluate these over a standard set of criteria to ensure appropriate models and methods can be selected for applied research.

\section{Acknowledgements}

This work was supported by the GenX Exposure study, on United States National Institutes of Health Superfund grant number P42 ES031009. The GenX Exposure Study is supported by research funding from the National Institute of Environmental Health Sciences (1R21ES029353), Center for Human Health and the Environment (CHHE) at NC State University (P30 ES025128), the Center for Environmental and Health Effects of PFAS (P42 ES0310095), and the NC Policy Collaboratory. This work was also supported by National Institutes of Health grants R01ES031651-03 and NIH R01ES031651-01, and National Science Foundation grant DMS2152887.

Code is posted at \href{GitHub}{https://github.com/nbwiecha/Chemical-Mixture-Methods-Comparison/tree/main}.

\begin{singlespace}
    \printbibliography

\end{singlespace}
\end{document}


\maketitle
\renewcommand\thesection{S\arabic{section}}
\renewcommand{\thetable}{S\arabic{table}}
\renewcommand{\thefigure}{S\arabic{figure}}

\section{Implementation of methods in the simulation study}
\subsection{Methods that test for individual components' association with the response}
For generalized additive models (GAMs) implemented in the R package \texttt{mgcv},\autocite{wood2011mgcv} variable selection using frequentist hypothesis tests, denoted in results tables and figures as ``GAM (HT)", as well as selection by variable selection penalties on the smooth terms, denoted in results tables and figures as ``GAM (VS)", was used. For fitting without variable selection penalties, the GAMs were estimated using 
{
restricted maximum likelihood (REML)} to allow adjustment for smoothing parameter uncertainty in hypothesis tests as recommended in \citeauthor{wood2017generalized}.\autocite{wood2017generalized} 

Bayesian kernel machine regression (BKMR),\autocite{bobb2015bayesian} implemented in the R package \texttt{bkmr},\autocite{bobb2018statistical} yields posterior inclusion probabilities (PIPs) for each variable.
PIP cutoffs of 0.5 and 0.95 were tried for determining whether a variable was selected into the model. The cutoff of 0.5 was tested because \citeauthor{lazarevic2020performance} found that a cutoff of 0.5 results in good variable selection performance.\autocite{lazarevic2020performance}
The cutoff of 0.95 was motivated by analogy to p-values and target Type I error probability of 0.05, however, PIPs are not designed to be analogous to p-values or have particular frequentist properties. 

For BKMR, 10,000 Markov Chain Monte Carlo (MCMC) iterations were used as burn-in and 10,000 iterations used as posterior samples. Package-default priors were used, and convergence of MCMC was checked using trace plots on example runs using simulation settings. 
Elastic net was implemented using the \texttt{glmnet} R package,\autocite{friedman2021package} with hyperparameters were chosen by 5-fold cross-validation (CV) performed by the \texttt{caret} package.\autocite{kuhn2007caret} Finally, a generalized linear model (GLM) was used, with identity link function and Gaussian likelihood, testing the null hypothesis $H_0: \beta_1=0$. GAMs did not include interaction terms for computational tractability.

\subsection{Methods that test for association between the overall mixture and the response}

For GAMs using frequentist hypothesis tests, to conduct a whole-mixture test of association, the individual tests' p-values were used with Bonferroni's correction.
{
An alternative approach is to conduct a generalized likelihood ratio test to compare the fit of an intercept-only GAM with that of the full GAM. This is noted in the documentation of \texttt{mgcv} to potentially be inaccurate, and in initial simulations did not control the Type I error rate with $p=10$ despite being based on frequentist justification, so was not included in the final results.} 
For BKMR, 
joint significance was tested by estimating a contrast $\gamma = h\bigl(Q_1(0.75), ..., Q_p(0.75)\bigr) - h\bigl(Q_1(0.25), ..., Q_p(0.25)\bigr)$, where $Q_j(w)$ is the $w$th quantile of exposure $j$, $w \in [0,1], j=1, ..., p$. An approximate test of $H_0$: $\gamma=0$ was performed at the $\alpha=0.05$ level by comparing $\frac{\hat{\gamma}}{\hat{\text{SE}(\gamma)}}$ to a $N(0,1)$ null distribution. This is an approximate form of a test recommended in \citeauthor{bobb2018statistical}.\autocite{bobb2018statistical} The hierarchical variable selection prior was also used to test joint association, with all exposure variables included in a single group.

For principal components regression (PCR), enough principal components were selected to explain at least $75\%$ of the exposure variance. An F-test on the estimated regression coefficients was performed to test for an effect of the mixture on the response. For weighted quantile sum (WQS) regression, implemented in the R package \texttt{gWQS},\autocite{gwqs} and quantile g-computation (QGC), implemented in the R package \texttt{gqcomp},\autocite{qgcomp} default settings were used. For WQS, 40\% of the sample was used to estimate weights and the remaining observations were used for inference, and a positive association between the mixture and response was assumed. 
As the GLM was used with a Gaussian likelihood, an F-test of joint significance was also used; with other likelihoods other tests of joint significance are available.

\section{Power curves}

This section provides all power curves for all scenarios and methods included.

\subsection{Individual component hypothesis tests' power curves}
\begin{figure}[H]
    \centering
    \includegraphics[width=0.9\linewidth]{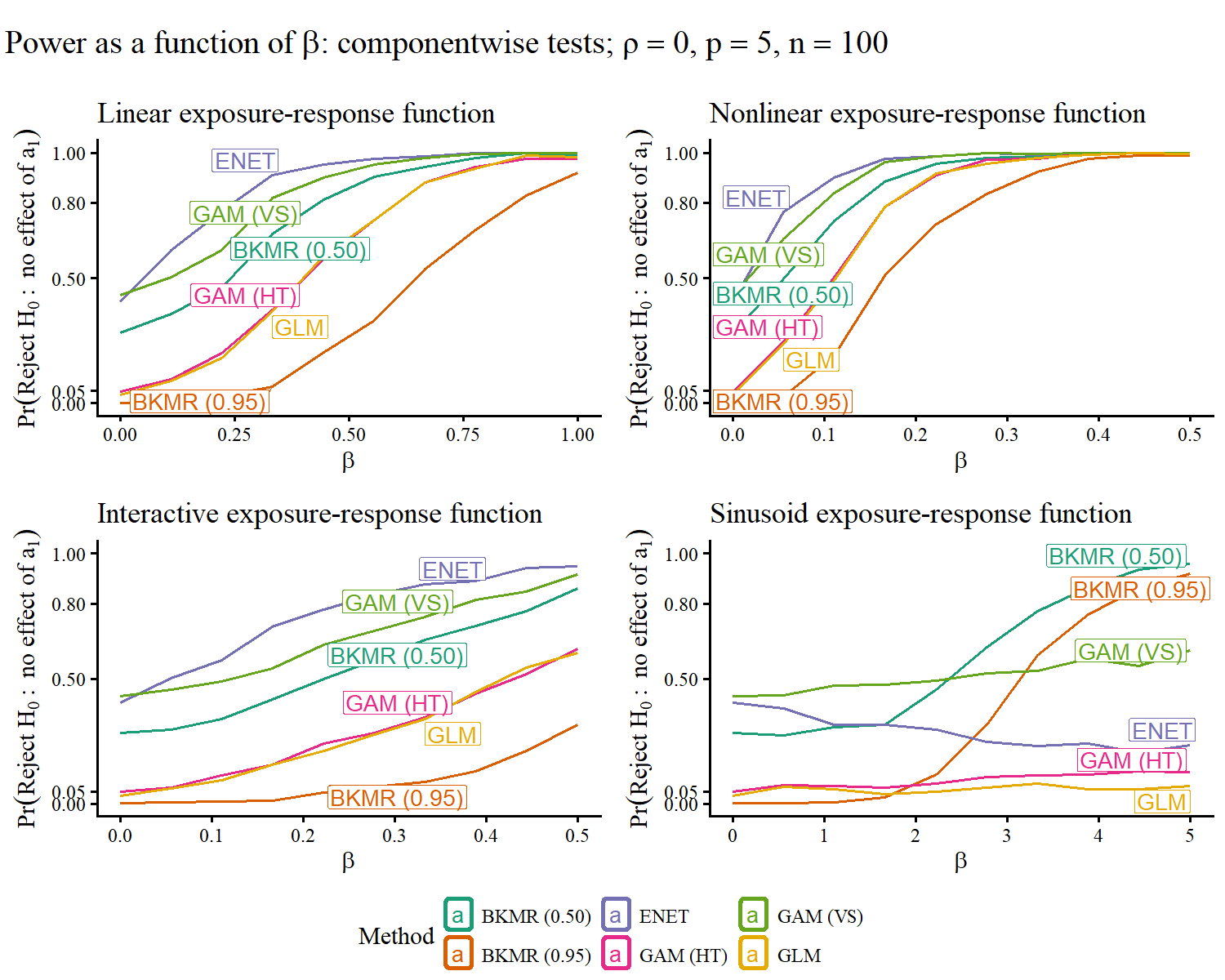}
    \caption{For each method (curve), the y-value is the probability of rejecting $H_0$: first mixture component $a_1$ not associated with the response, for different strengths of association $\beta$ (x-axis). When $\beta=0$, the value of the curve is the estimated Type I error probability. The targeted Type I error rate was $0.05$. When $\beta \neq 0$, the value of the curve is the power of the hypothesis test. Better methods have Type I error rate at most 0.05, and power curve above other methods. Exposure-response functions used are linear, nonlinear, linear interaction, and sinusoid (nonlinear interaction).}
    \label{fig:power_varsel_n100_p5_rho0}
\end{figure}

\begin{figure}[H]
    \centering
    \includegraphics[width=0.9\linewidth]{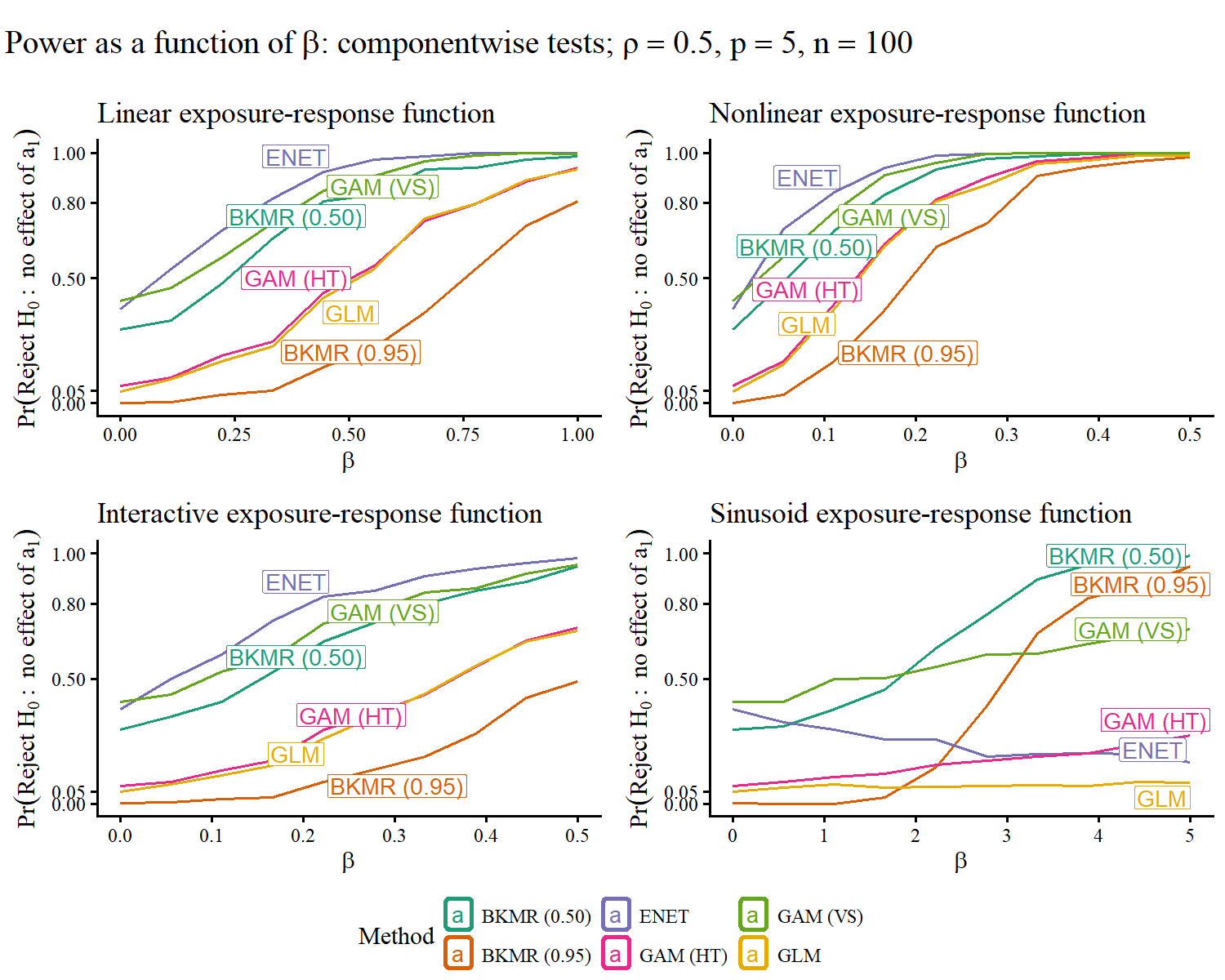}
    \caption{For each method (curve), the y-value is the probability of rejecting $H_0$: first mixture component $a_1$ not associated with the response, for different strengths of association $\beta$ (x-axis). When $\beta=0$, the value of the curve is the estimated Type I error probability. The targeted Type I error rate was $0.05$. When $\beta \neq 0$, the value of the curve is the power of the hypothesis test. Better methods have Type I error rate at most 0.05, and power curve above other methods. Exposure-response functions used are linear, nonlinear, linear interaction, and sinusoid (nonlinear interaction).}
    \label{fig:power_varsel_n100_p5_rho5}
\end{figure}

\begin{figure}[H]
    \centering
    \includegraphics[width=0.9\linewidth]{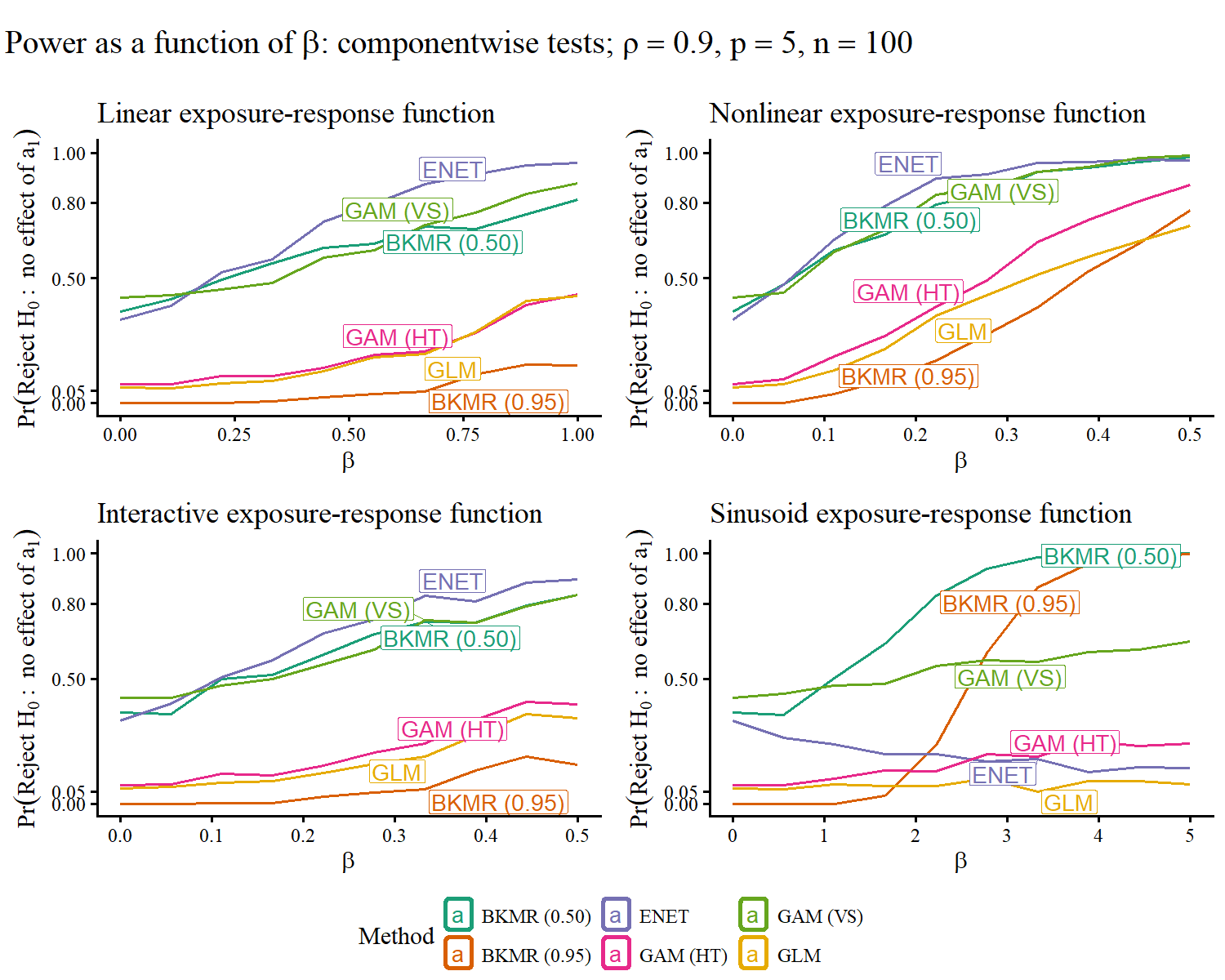}
    \caption{For each method (curve), the y-value is the probability of rejecting $H_0$: first mixture component $a_1$ not associated with the response, for different strengths of association $\beta$ (x-axis). When $\beta=0$, the value of the curve is the estimated Type I error probability. The targeted Type I error rate was $0.05$. When $\beta \neq 0$, the value of the curve is the power of the hypothesis test. Better methods have Type I error rate at most 0.05, and power curve above other methods. Exposure-response functions used are linear, nonlinear, linear interaction, and sinusoid (nonlinear interaction).}
    \label{fig:power_varsel_n100_p5_rho9}
\end{figure}

\begin{figure}[H]
    \centering
    \includegraphics[width=0.9\linewidth]{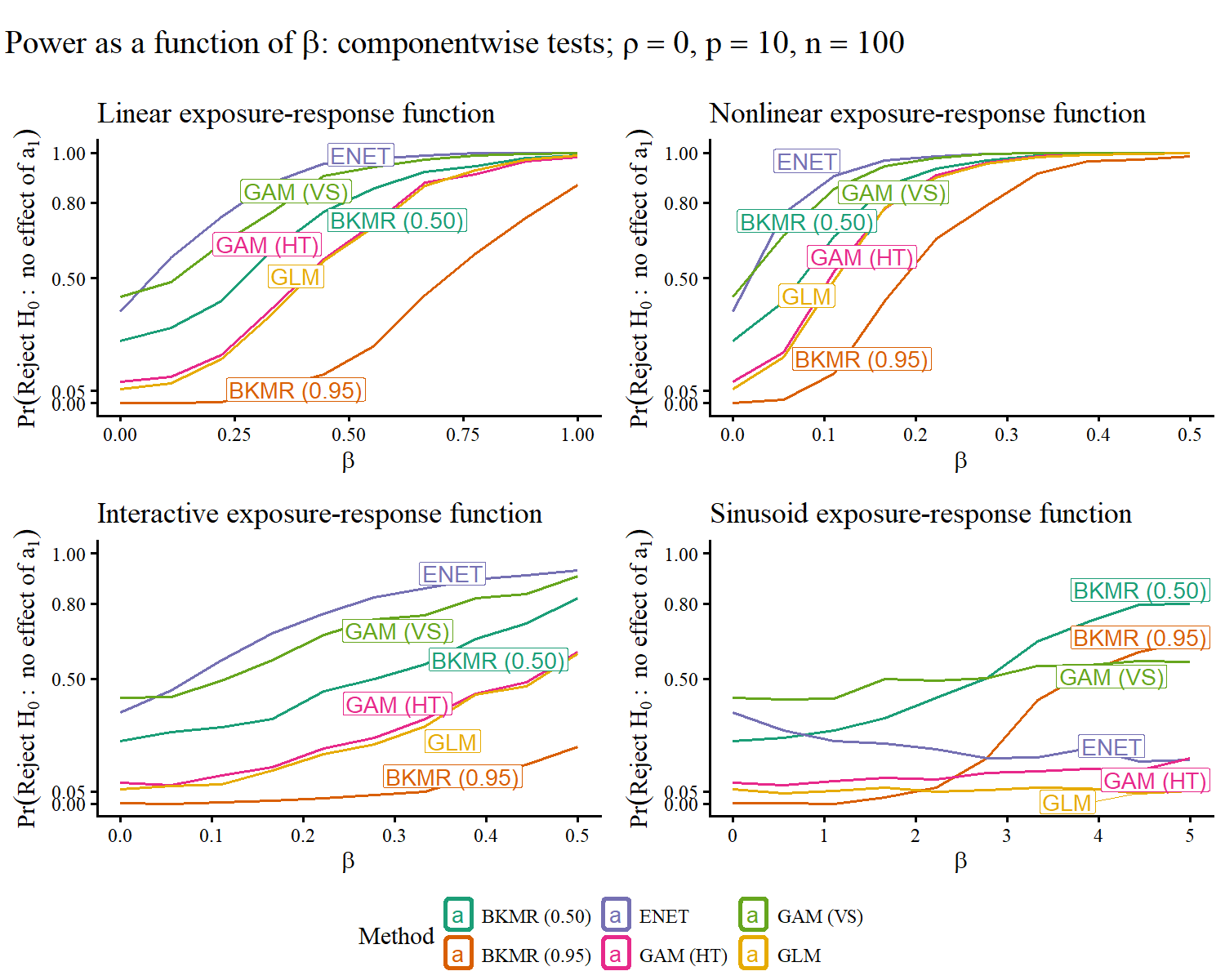}
    \caption{For each method (curve), the y-value is the probability of rejecting $H_0$: first mixture component $a_1$ not associated with the response, for different strengths of association $\beta$ (x-axis). When $\beta=0$, the value of the curve is the estimated Type I error probability. The targeted Type I error rate was $0.05$. When $\beta \neq 0$, the value of the curve is the power of the hypothesis test. Better methods have Type I error rate at most 0.05, and power curve above other methods. Exposure-response functions used are linear, nonlinear, linear interaction, and sinusoid (nonlinear interaction).}
    \label{fig:power_varsel_n100_p10_rho0}
\end{figure}

\begin{figure}[H]
    \centering
    \includegraphics[width=0.9\linewidth]{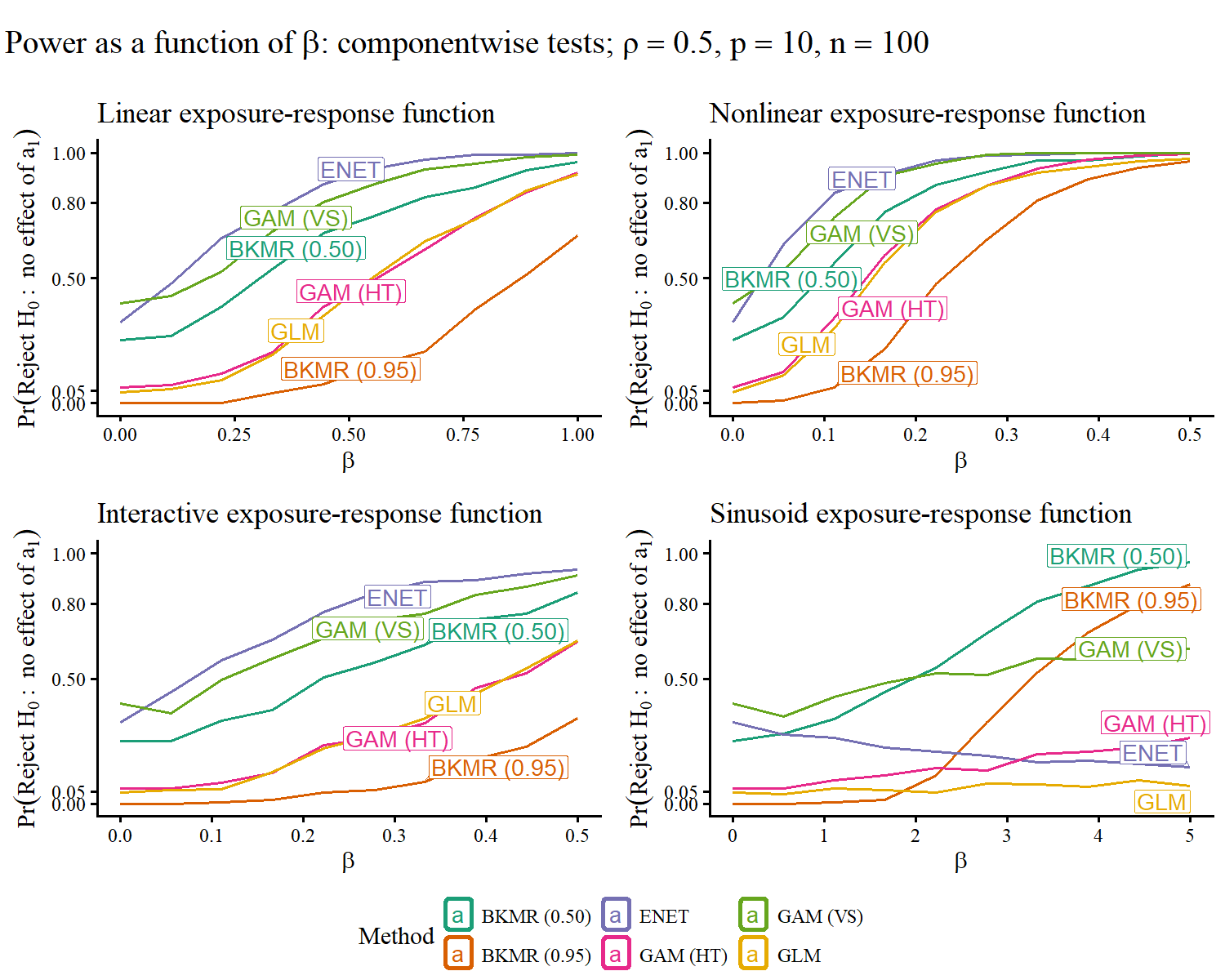}
    \caption{For each method (curve), the y-value is the probability of rejecting $H_0$: first mixture component $a_1$ not associated with the response, for different strengths of association $\beta$ (x-axis). When $\beta=0$, the value of the curve is the estimated Type I error probability. The targeted Type I error rate was $0.05$. When $\beta \neq 0$, the value of the curve is the power of the hypothesis test. Better methods have Type I error rate at most 0.05, and power curve above other methods. Exposure-response functions used are linear, nonlinear, linear interaction, and sinusoid (nonlinear interaction).}
    \label{fig:power_varsel_n100_p10_rho5}
\end{figure}

\begin{figure}[H]
    \centering
    \includegraphics[width=0.9\linewidth]{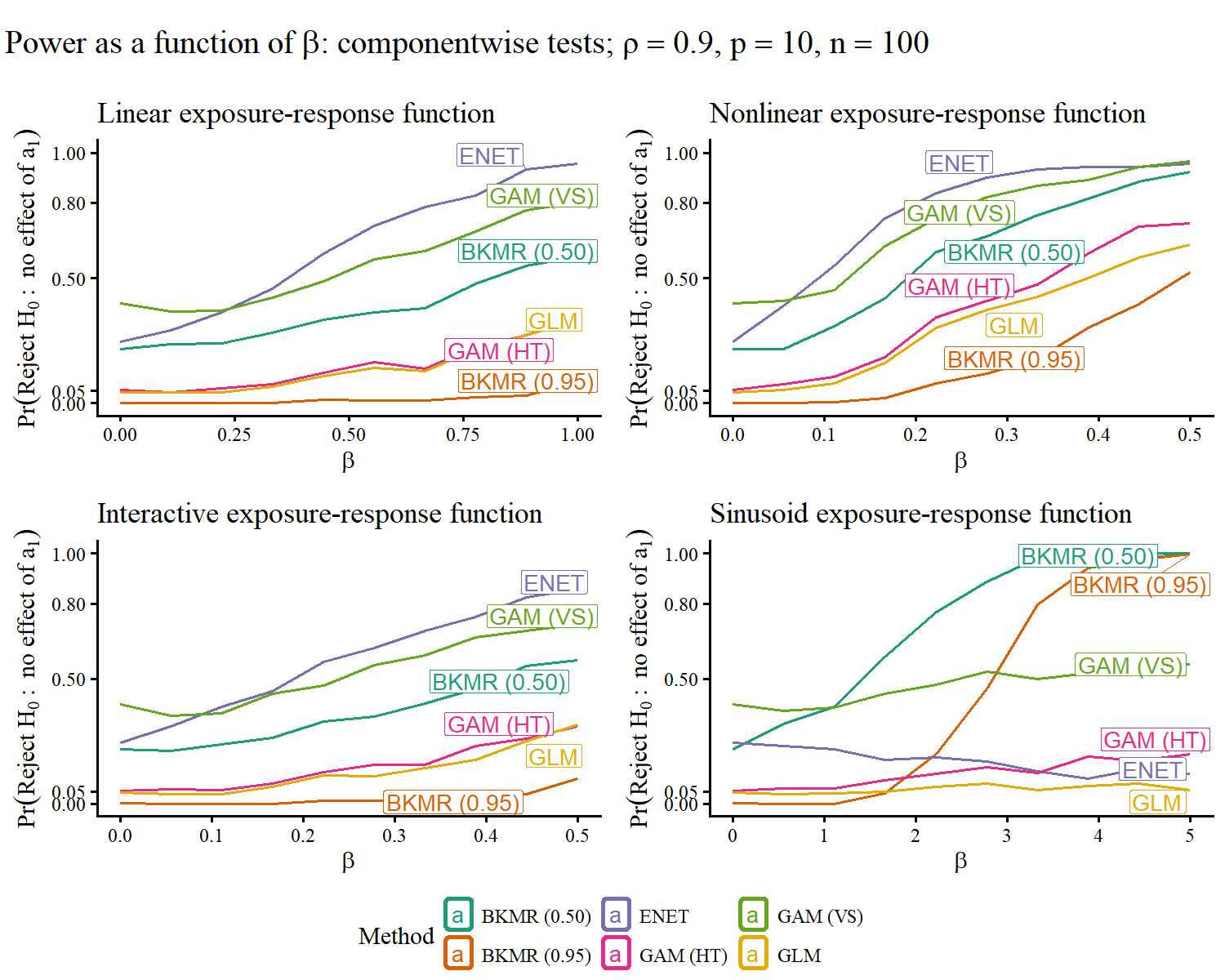}
    \caption{For each method (curve), the y-value is the probability of rejecting $H_0$: first mixture component $a_1$ not associated with the response, for different strengths of association $\beta$ (x-axis). When $\beta=0$, the value of the curve is the estimated Type I error probability. The targeted Type I error rate was $0.05$. When $\beta \neq 0$, the value of the curve is the power of the hypothesis test. Better methods have Type I error rate at most 0.05, and power curve above other methods. Exposure-response functions used are linear, nonlinear, linear interaction, and sinusoid (nonlinear interaction).}
    \label{fig:power_varsel_n100_p10_rho9}
\end{figure}

\begin{figure}[H]
    \centering
    \includegraphics[width=0.9\linewidth]{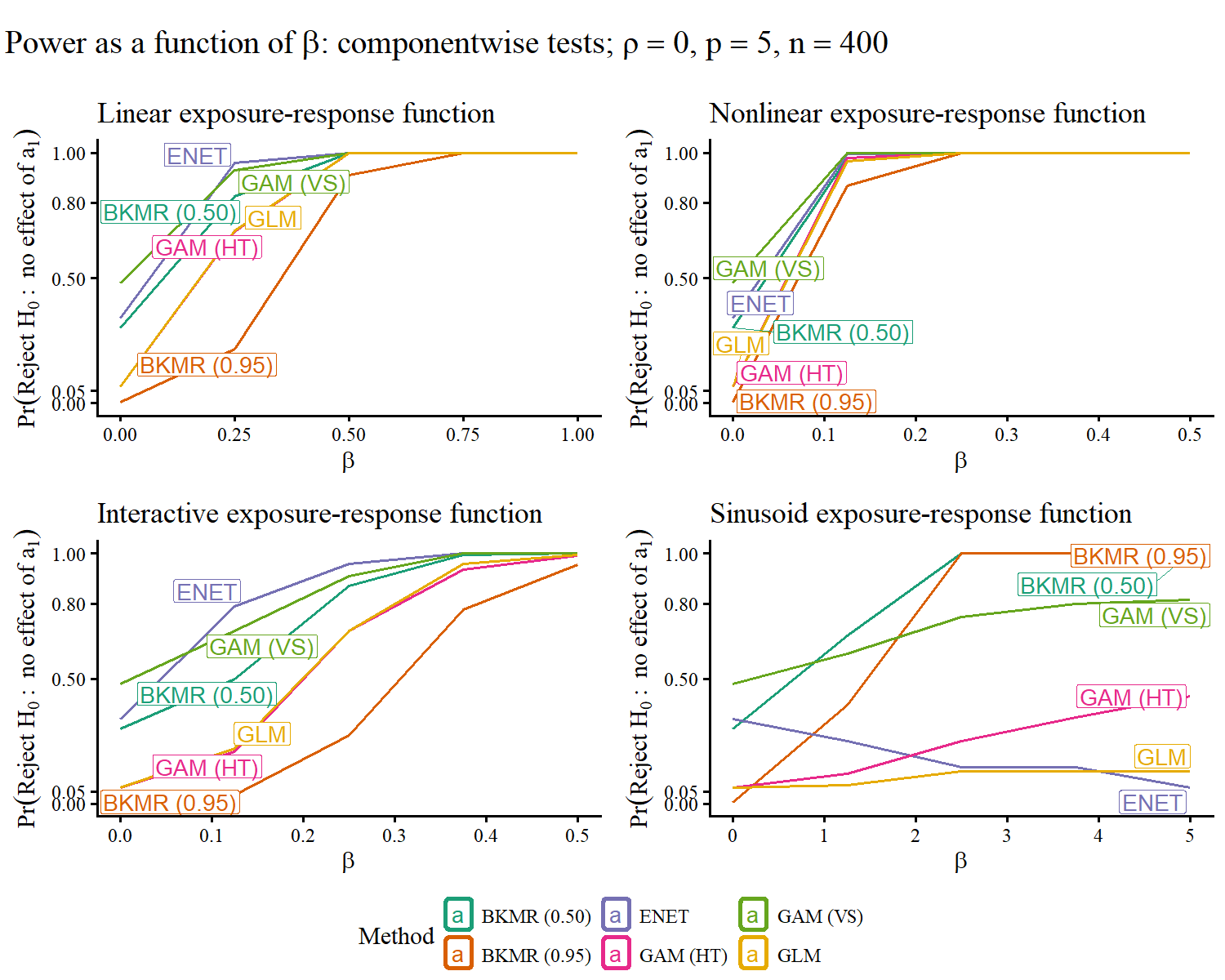}
    \caption{For each method (curve), the y-value is the probability of rejecting $H_0$: first mixture component $a_1$ not associated with the response, for different strengths of association $\beta$ (x-axis). When $\beta=0$, the value of the curve is the estimated Type I error probability. The targeted Type I error rate was $0.05$. When $\beta \neq 0$, the value of the curve is the power of the hypothesis test. Better methods have Type I error rate at most 0.05, and power curve above other methods. Exposure-response functions used are linear, nonlinear, linear interaction, and sinusoid (nonlinear interaction).}
    \label{fig:power_varsel_n400_p5_rho0}
\end{figure}

\begin{figure}[H]
    \centering
    \includegraphics[width=0.9\linewidth]{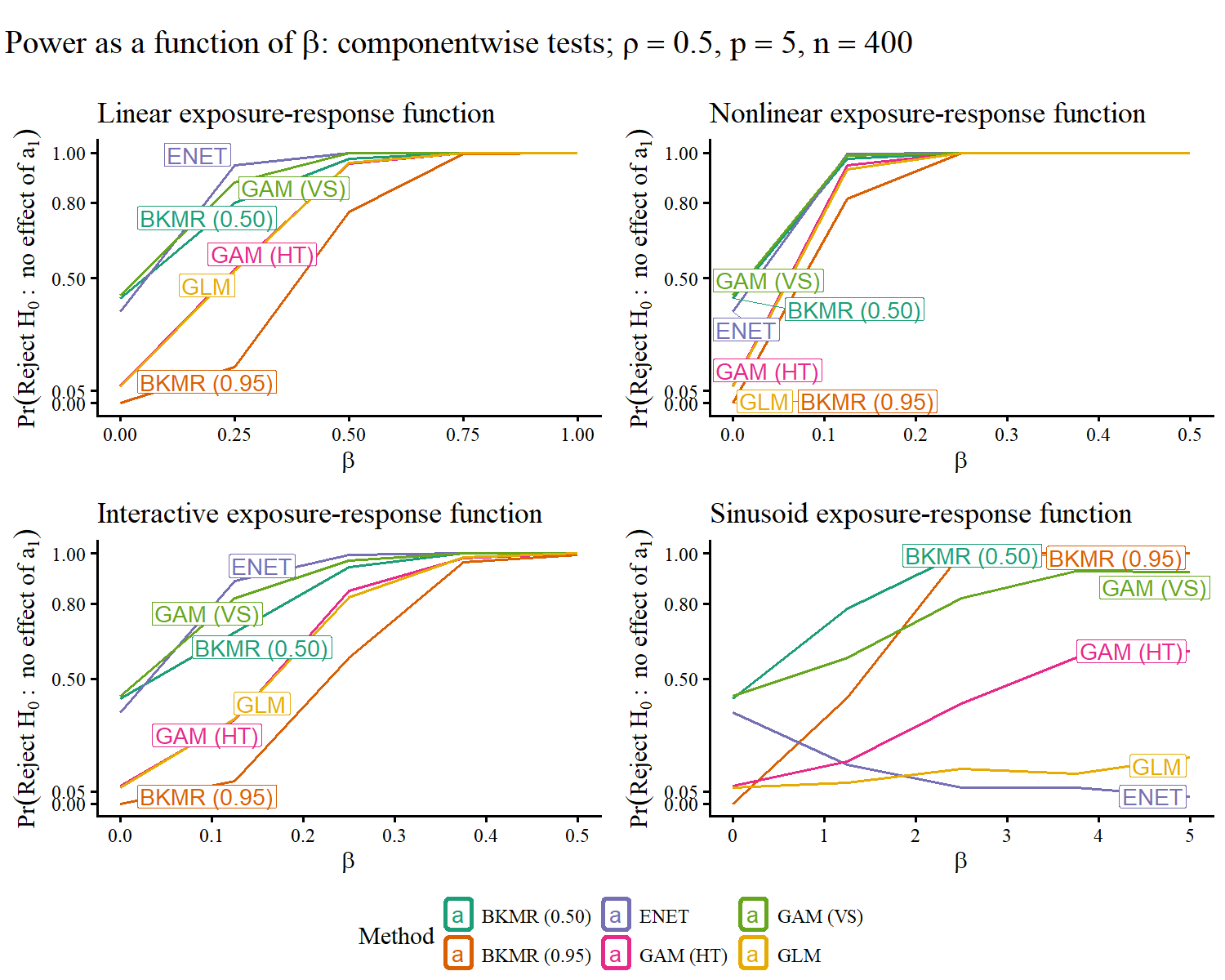}
    \caption{For each method (curve), the y-value is the probability of rejecting $H_0$: first mixture component $a_1$ not associated with the response, for different strengths of association $\beta$ (x-axis). When $\beta=0$, the value of the curve is the estimated Type I error probability. The targeted Type I error rate was $0.05$. When $\beta \neq 0$, the value of the curve is the power of the hypothesis test. Better methods have Type I error rate at most 0.05, and power curve above other methods. Exposure-response functions used are linear, nonlinear, linear interaction, and sinusoid (nonlinear interaction).}
    \label{fig:power_varsel_n400_p5_rho5}
\end{figure}

\subsection{Whole-mixture hypothesis tests' power curves}

\begin{figure}[H]
    \centering
    \includegraphics[width=0.9\linewidth]{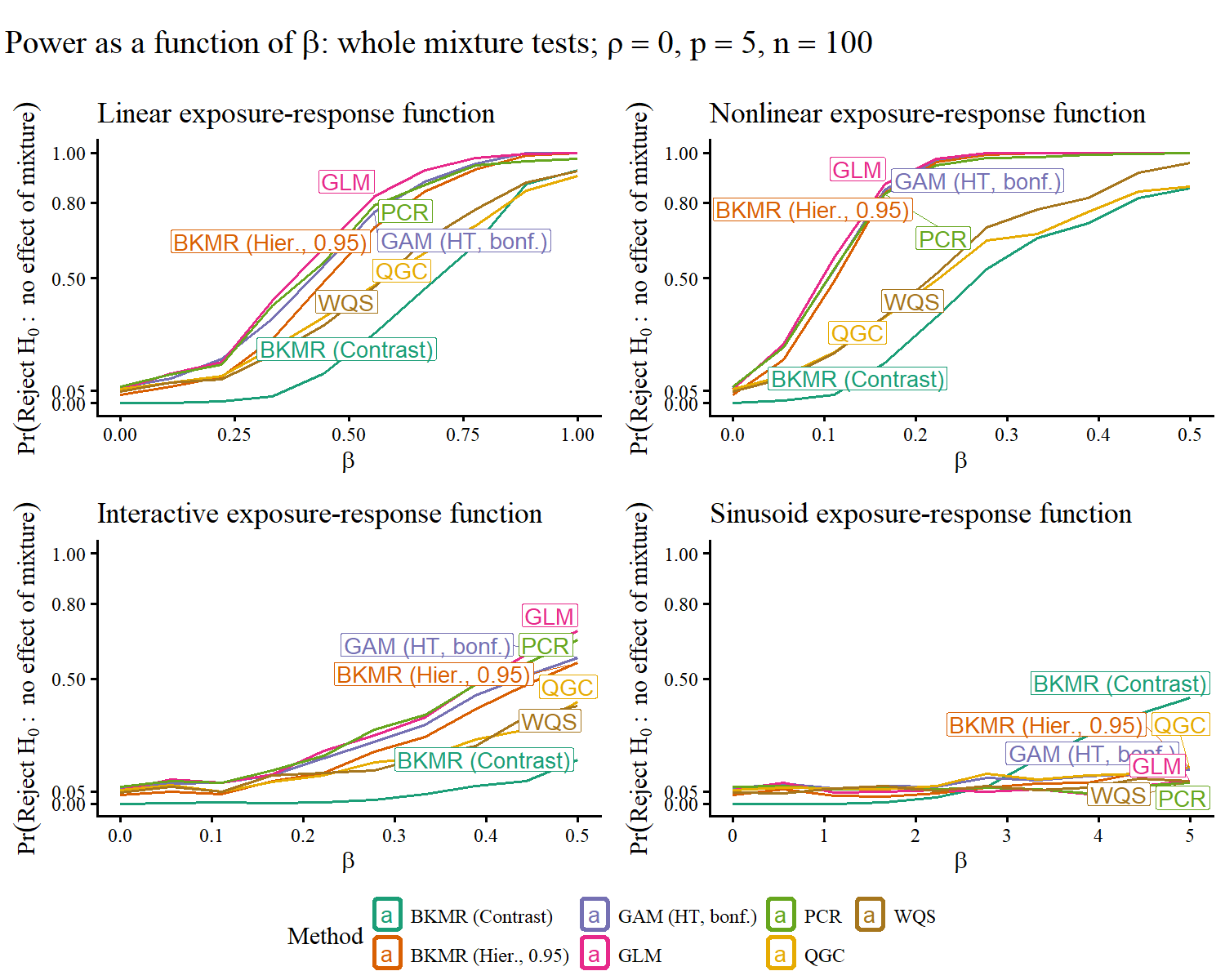}
    \caption{For each method (curve), the y-value is the probability of rejecting $H_0$: the overall mixture is not associated with the response, for different strengths of association $\beta$ (x-axis). When $\beta=0$, the value of the curve is the estimated Type I error probability. The targeted Type I error rate was $0.05$. When $\beta \neq 0$, the value of the curve is the power of the hypothesis test. Better methods have Type I error rate at most 0.05, and power curve above other methods. Exposure-response functions used are linear, nonlinear, linear interaction, and sinusoid (nonlinear interaction).}
    \label{fig:power_idx_n100_p5_rho0}
\end{figure}

\begin{figure}[H]
    \centering
    \includegraphics[width=0.9\linewidth]{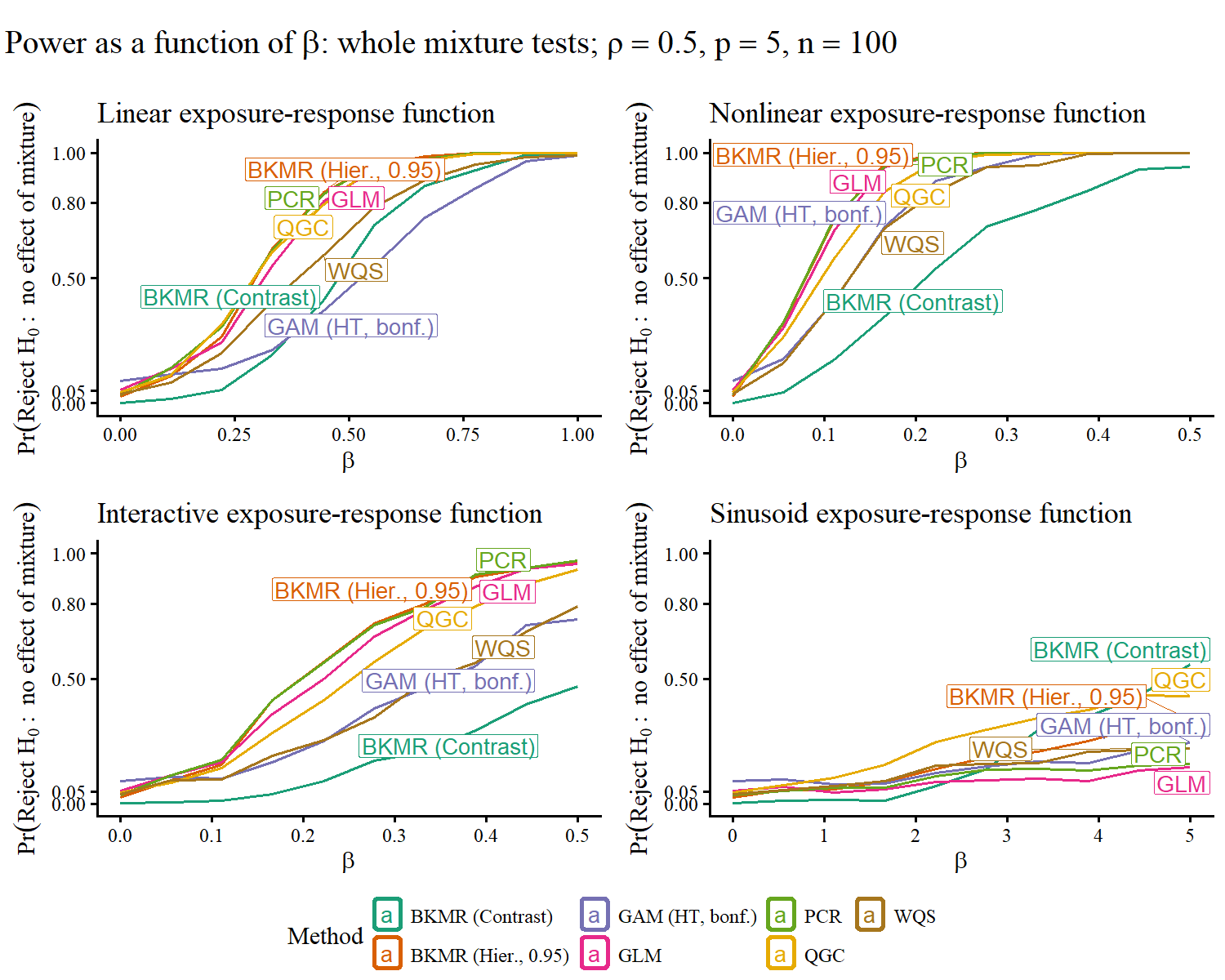}
    \caption{For each method (curve), the y-value is the probability of rejecting $H_0$: the overall mixture is not associated with the response, for different strengths of association $\beta$ (x-axis). When $\beta=0$, the value of the curve is the estimated Type I error probability. The targeted Type I error rate was $0.05$. When $\beta \neq 0$, the value of the curve is the power of the hypothesis test. Better methods have Type I error rate at most 0.05, and power curve above other methods. Exposure-response functions used are linear, nonlinear, linear interaction, and sinusoid (nonlinear interaction).}
    \label{fig:power_idx_n100_p5_rho5}
\end{figure}

\begin{figure}[H]
    \centering
    \includegraphics[width=0.9\linewidth]{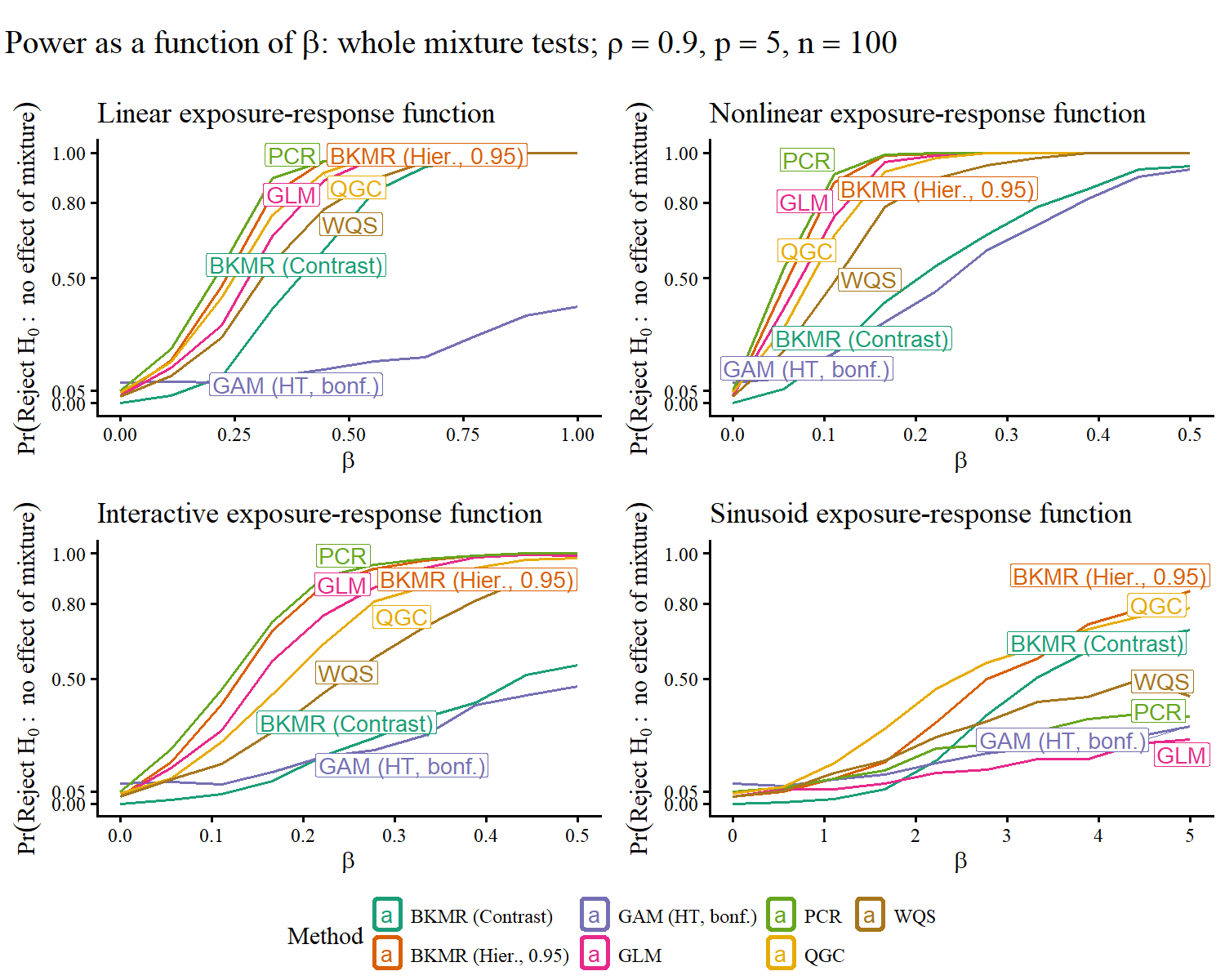}
    \caption{For each method (curve), the y-value is the probability of rejecting $H_0$: the overall mixture is not associated with the response, for different strengths of association $\beta$ (x-axis). When $\beta=0$, the value of the curve is the estimated Type I error probability. The targeted Type I error rate was $0.05$. When $\beta \neq 0$, the value of the curve is the power of the hypothesis test. Better methods have Type I error rate at most 0.05, and power curve above other methods. Exposure-response functions used are linear, nonlinear, linear interaction, and sinusoid (nonlinear interaction).}
    \label{fig:power_idx_n100_p5_rho9}
\end{figure}

\begin{figure}[H]
    \centering
    \includegraphics[width=0.9\linewidth]{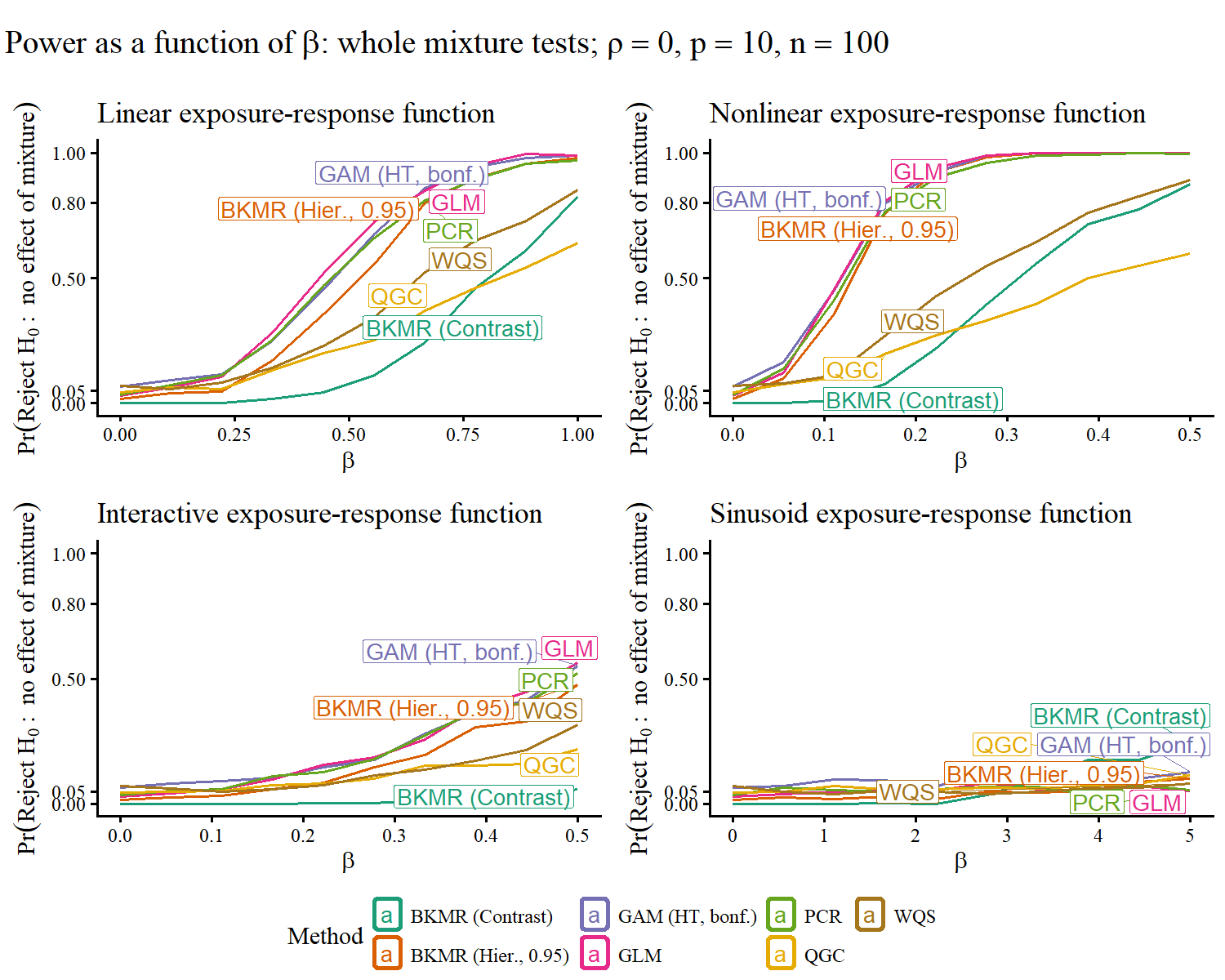}
    \caption{For each method (curve), the y-value is the probability of rejecting $H_0$: the overall mixture is not associated with the response, for different strengths of association $\beta$ (x-axis). When $\beta=0$, the value of the curve is the estimated Type I error probability. The targeted Type I error rate was $0.05$. When $\beta \neq 0$, the value of the curve is the power of the hypothesis test. Better methods have Type I error rate at most 0.05, and power curve above other methods. Exposure-response functions used are linear, nonlinear, linear interaction, and sinusoid (nonlinear interaction).}
    \label{fig:power_idx_n100_p10_rho0}
\end{figure}

\begin{figure}[H]
    \centering
    \includegraphics[width=0.9\linewidth]{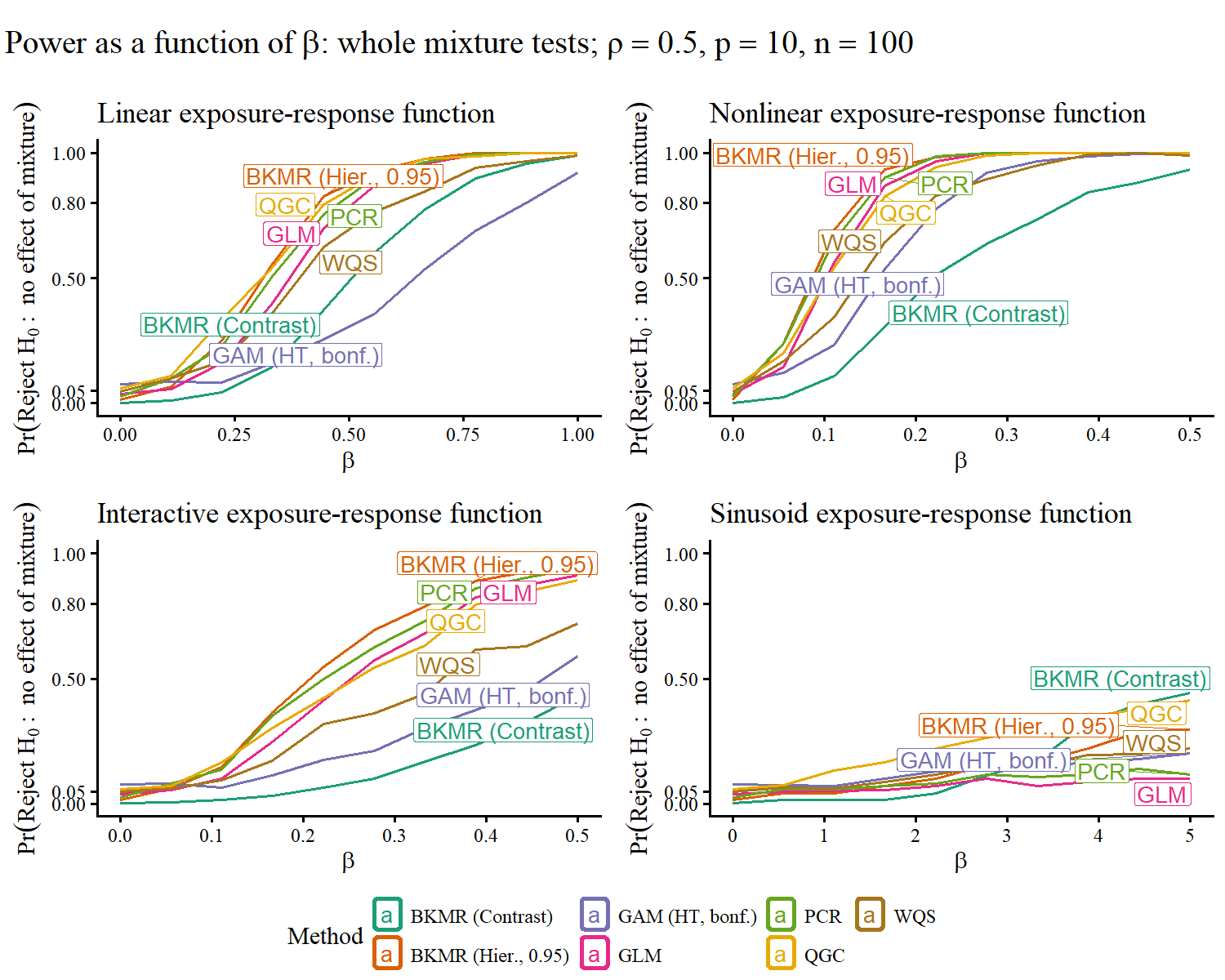}
    \caption{For each method (curve), the y-value is the probability of rejecting $H_0$: the overall mixture is not associated with the response, for different strengths of association $\beta$ (x-axis). When $\beta=0$, the value of the curve is the estimated Type I error probability. The targeted Type I error rate was $0.05$. When $\beta \neq 0$, the value of the curve is the power of the hypothesis test. Better methods have Type I error rate at most 0.05, and power curve above other methods. Exposure-response functions used are linear, nonlinear, linear interaction, and sinusoid (nonlinear interaction).}
    \label{fig:power_idx_n100_p10_rho5}
\end{figure}

\begin{figure}[H]
    \centering
    \includegraphics[width=0.9\linewidth]{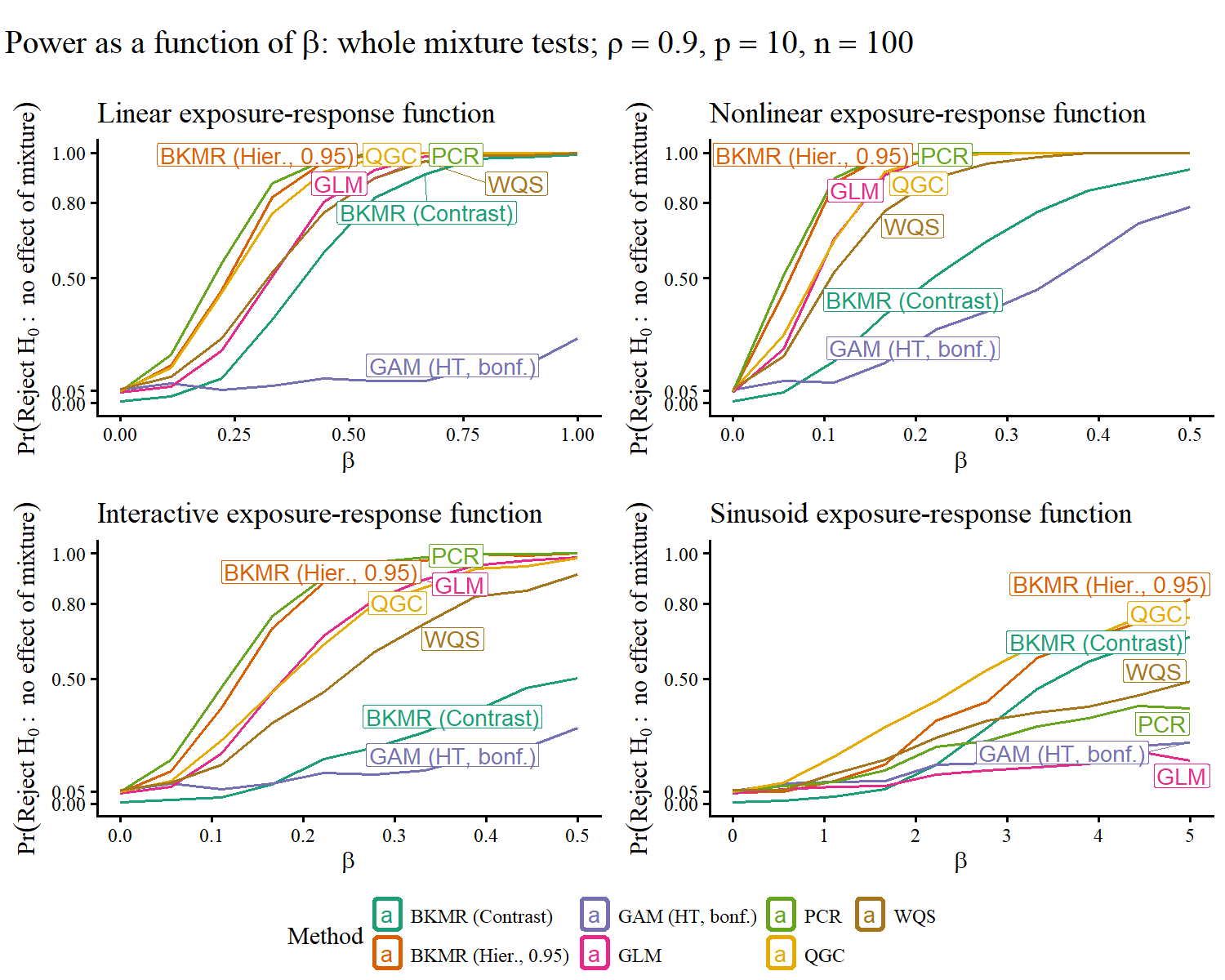}
    \caption{For each method (curve), the y-value is the probability of rejecting $H_0$: the overall mixture is not associated with the response, for different strengths of association $\beta$ (x-axis). When $\beta=0$, the value of the curve is the estimated Type I error probability. The targeted Type I error rate was $0.05$. When $\beta \neq 0$, the value of the curve is the power of the hypothesis test. Better methods have Type I error rate at most 0.05, and power curve above other methods. Exposure-response functions used are linear, nonlinear, linear interaction, and sinusoid (nonlinear interaction).}
    \label{fig:power_idx_n100_p10_rho9}
\end{figure}

\begin{figure}[H]
    \centering
    \includegraphics[width=0.5\linewidth]{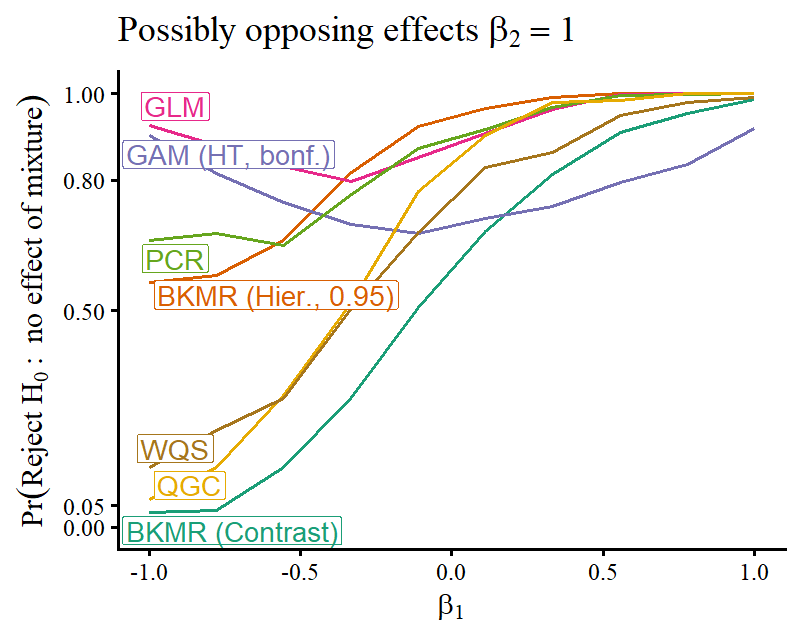}
    \caption{Power curves for whole-mixture hypothesis tests for $n=100, p=10, \rho=0.5$, possibly opposing effects. For each method (curve), the y-value is the probability of rejecting $H_0$: the overall mixture is not associated with the response, for different strengths of association $\beta$ (x-axis). Exposure-response function used is linear with possibly opposing effects of mixture components. Note that the x-axis endpoints have changed as $E(y_i|a_{1i}, a_{2i}) = \beta_1a_1 + \beta_2 a_2$, and $\beta_2$ is fixed at 1, and $\beta_1$ now varies from $-1$ to $1$. Therefore better methods have all values of their power curve above other methods'.}
    \label{fig:power_idx_n100_p10_rho5_opp}
\end{figure}

\begin{figure}[H]
    \centering
    \includegraphics[width=0.9\linewidth]{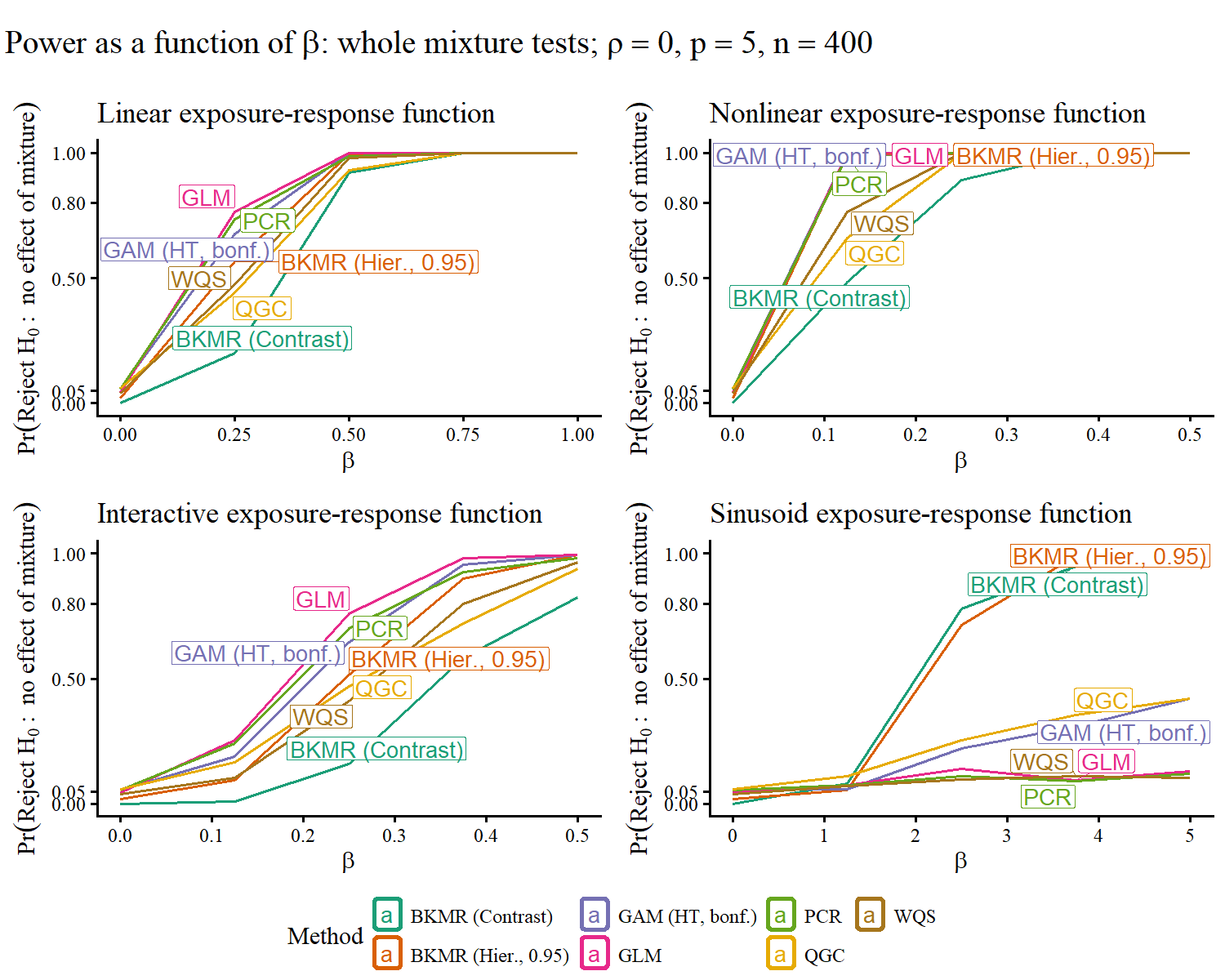}
    \caption{For each method (curve), the y-value is the probability of rejecting $H_0$: the overall mixture is not associated with the response, for different strengths of association $\beta$ (x-axis). When $\beta=0$, the value of the curve is the estimated Type I error probability. The targeted Type I error rate was $0.05$. When $\beta \neq 0$, the value of the curve is the power of the hypothesis test. Better methods have Type I error rate at most 0.05, and power curve above other methods. Exposure-response functions used are linear, nonlinear, linear interaction, and sinusoid (nonlinear interaction).}
    \label{fig:power_idx_n400_p5_rho0}
\end{figure}

\begin{figure}[H]
    \centering
    \includegraphics[width=0.9\linewidth]{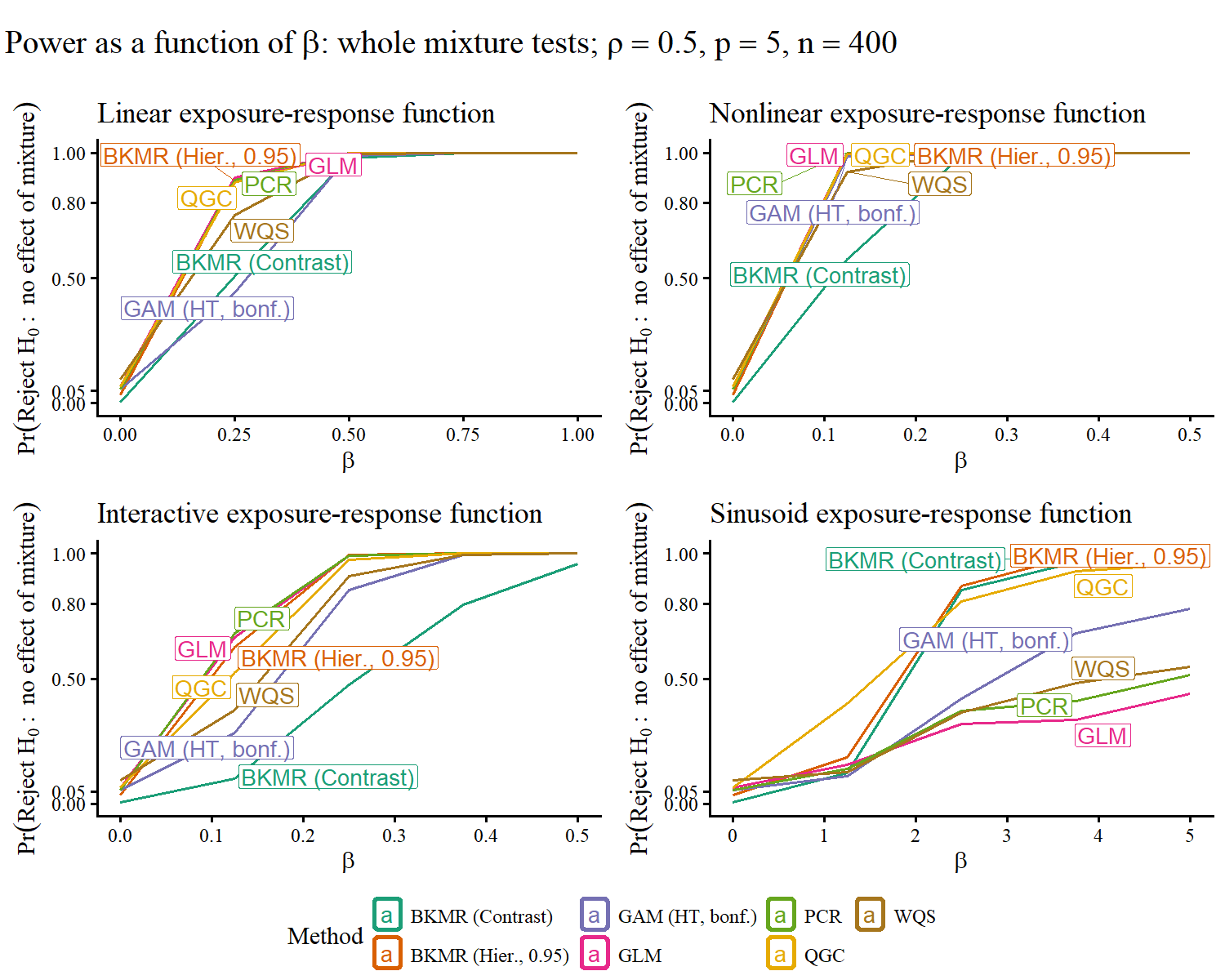}
    \caption{For each method (curve), the y-value is the probability of rejecting $H_0$: the overall mixture is not associated with the response, for different strengths of association $\beta$ (x-axis). When $\beta=0$, the value of the curve is the estimated Type I error probability. The targeted Type I error rate was $0.05$. When $\beta \neq 0$, the value of the curve is the power of the hypothesis test. Better methods have Type I error rate at most 0.05, and power curve above other methods. Exposure-response functions used are linear, nonlinear, linear interaction, and sinusoid (nonlinear interaction).}
    \label{fig:power_idx_n400_p5_rho5}
\end{figure}

\begin{figure}[H]
    \centering
    \includegraphics[width=0.5\linewidth]{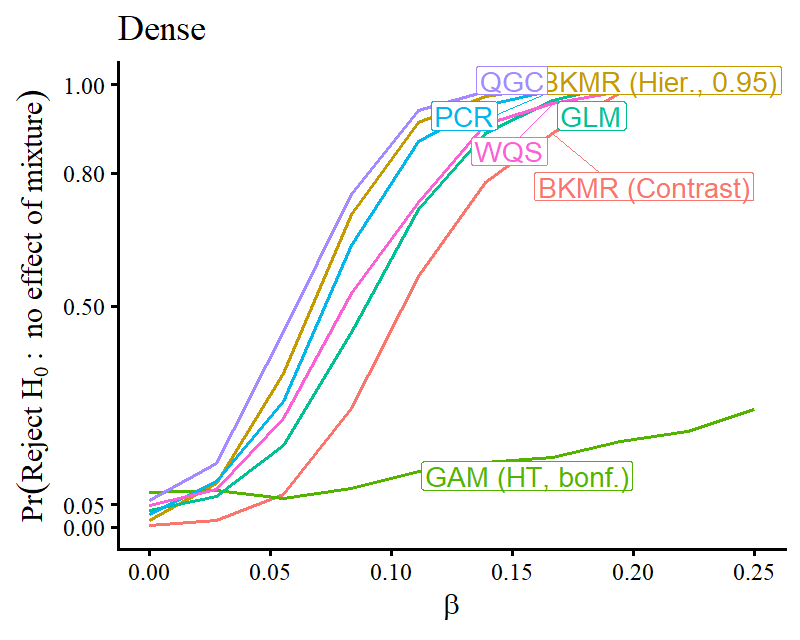}
    \caption{{
    For each method (curve), the y-value is the probability of rejecting $H_0$: the overall mixture is not associated with the response, for different strengths of association $\beta$ (x-axis). When $\beta=0$, the value of the curve is the estimated Type I error probability. The targeted Type I error rate was $0.05$. When $\beta \neq 0$, the value of the curve is the power of the hypothesis test. Better methods have Type I error rate at most 0.05, and power curve above other methods.} The exposure-response function used is dense (all exposures affecting the response) with $p=10$ exposures.}
    \label{fig:power_idx_p5_rho5_dense}
\end{figure}

\subsection{Prediction error plots}

\begin{figure}[H]
    \centering
    \includegraphics[width=0.9\linewidth]{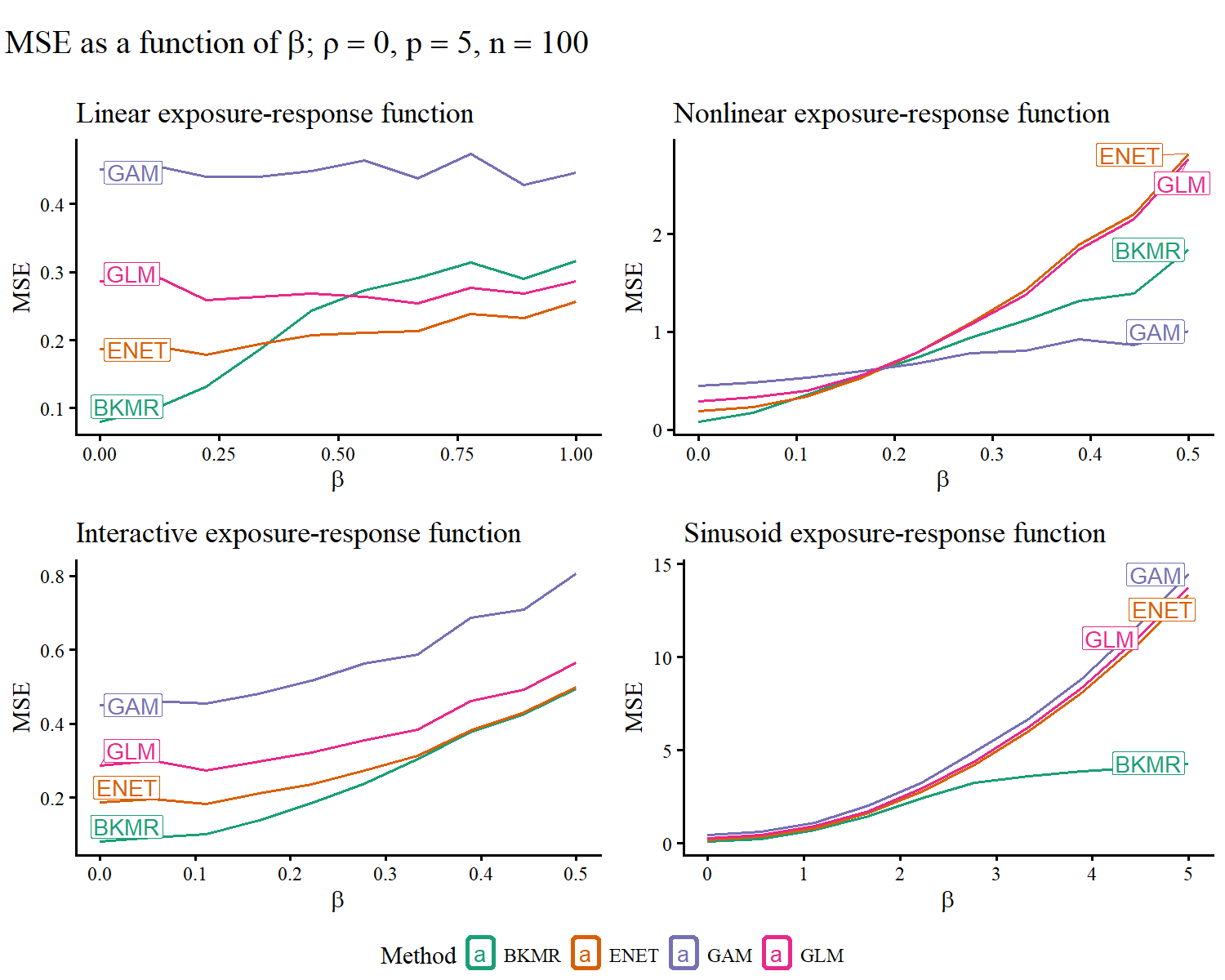}
    \caption{For each method (curve), the y-value is the prediction mean squared error (MSE) on new data, for different strengths of association $\beta$ (x-axis). Better methods have lower MSE curves. The new data are generated from the same exposure-response functions as the data used to fit the model, but have no random variation added. Exposure-response functions used are linear, nonlinear, linear interaction, and sinusoid (nonlinear interaction).}
    \label{fig:mse_n100_p5_rho0}
\end{figure}

\begin{figure}[H]
    \centering
    \includegraphics[width=0.9\linewidth]{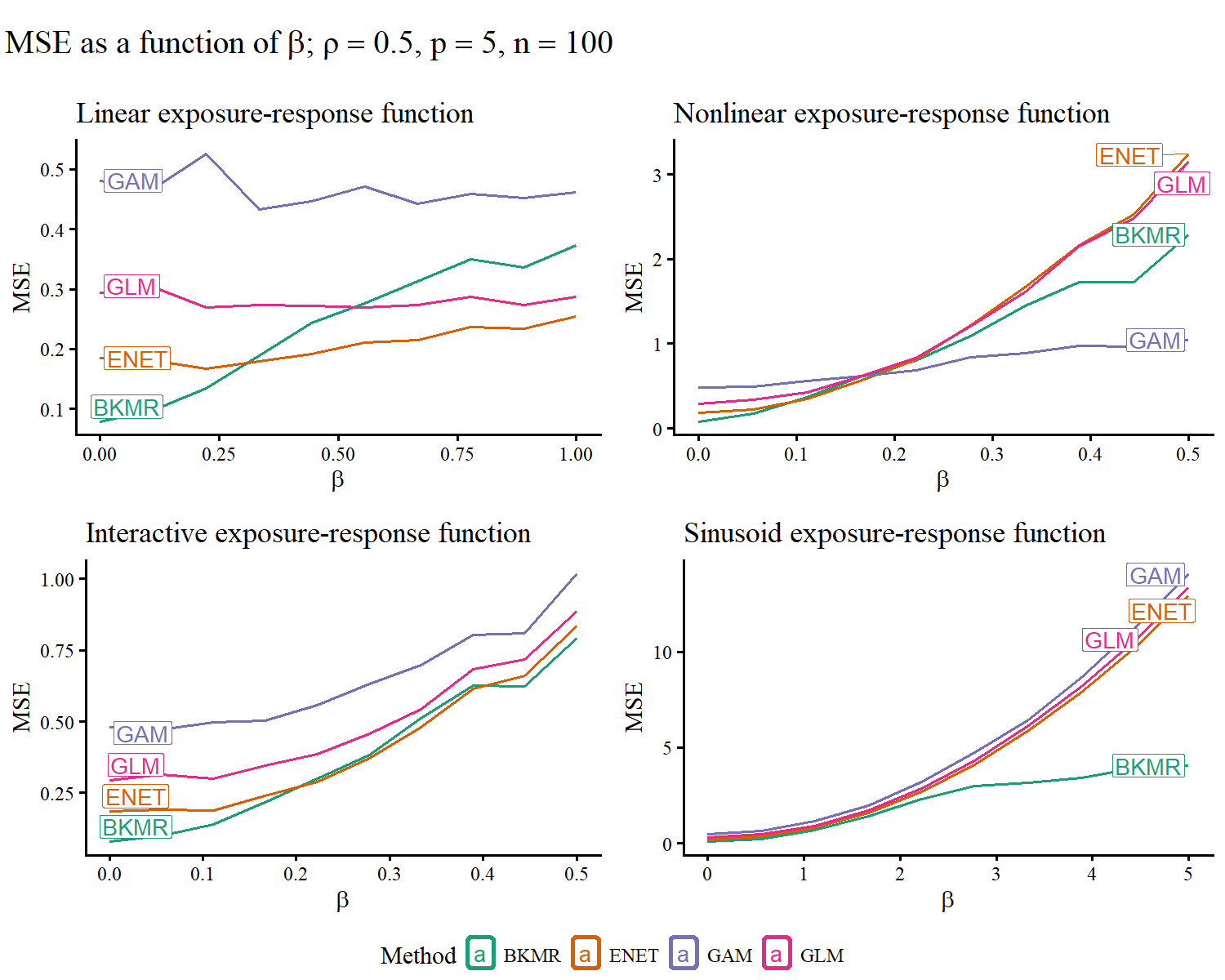}
    \caption{For each method (curve), the y-value is the prediction mean squared error (MSE) on new data, for different strengths of association $\beta$ (x-axis). Better methods have lower MSE curves. The new data are generated from the same exposure-response functions as the data used to fit the model, but have no random variation added. Exposure-response functions used are linear, nonlinear, linear interaction, and sinusoid (nonlinear interaction).}
    \label{fig:mse_n100_p5_rho5}
\end{figure}

\begin{figure}[H]
    \centering
    \includegraphics[width=0.9\linewidth]{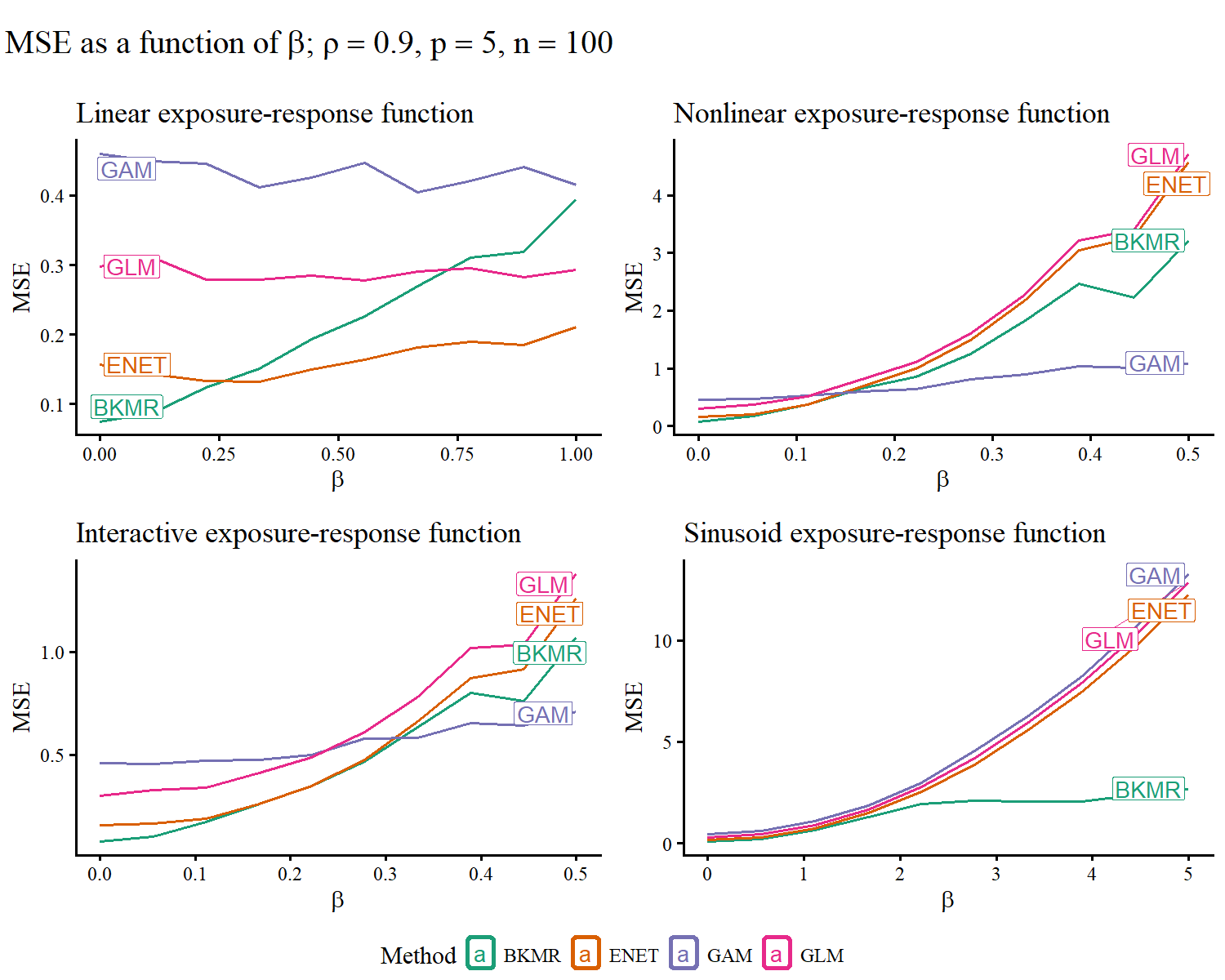}
    \caption{For each method (curve), the y-value is the prediction mean squared error (MSE) on new data, for different strengths of association $\beta$ (x-axis). Better methods have lower MSE curves. The new data are generated from the same exposure-response functions as the data used to fit the model, but have no random variation added. Exposure-response functions used are linear, nonlinear, linear interaction, and sinusoid (nonlinear interaction).}
    \label{fig:mse_n100_p5_rho9}
\end{figure}

\begin{figure}[H]
    \centering
    \includegraphics[width=0.9\linewidth]{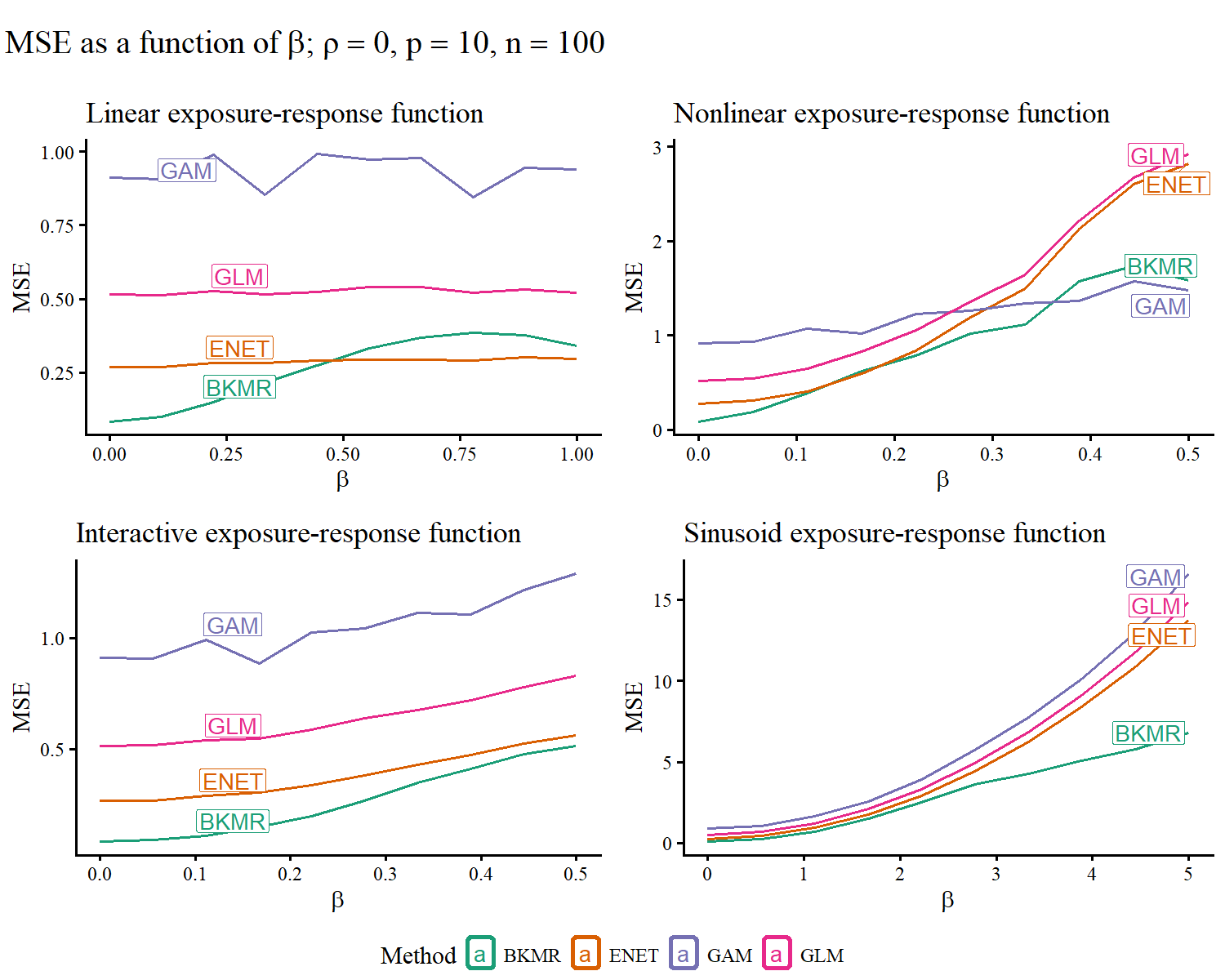}
    \caption{For each method (curve), the y-value is the prediction mean squared error (MSE) on new data, for different strengths of association $\beta$ (x-axis). Better methods have lower MSE curves. The new data are generated from the same exposure-response functions as the data used to fit the model, but have no random variation added. Exposure-response functions used are linear, nonlinear, linear interaction, and sinusoid (nonlinear interaction).}
    \label{fig:mse_n100_p10_rho0}
\end{figure}

\begin{figure}[H]
    \centering
    \includegraphics[width=0.9\linewidth]{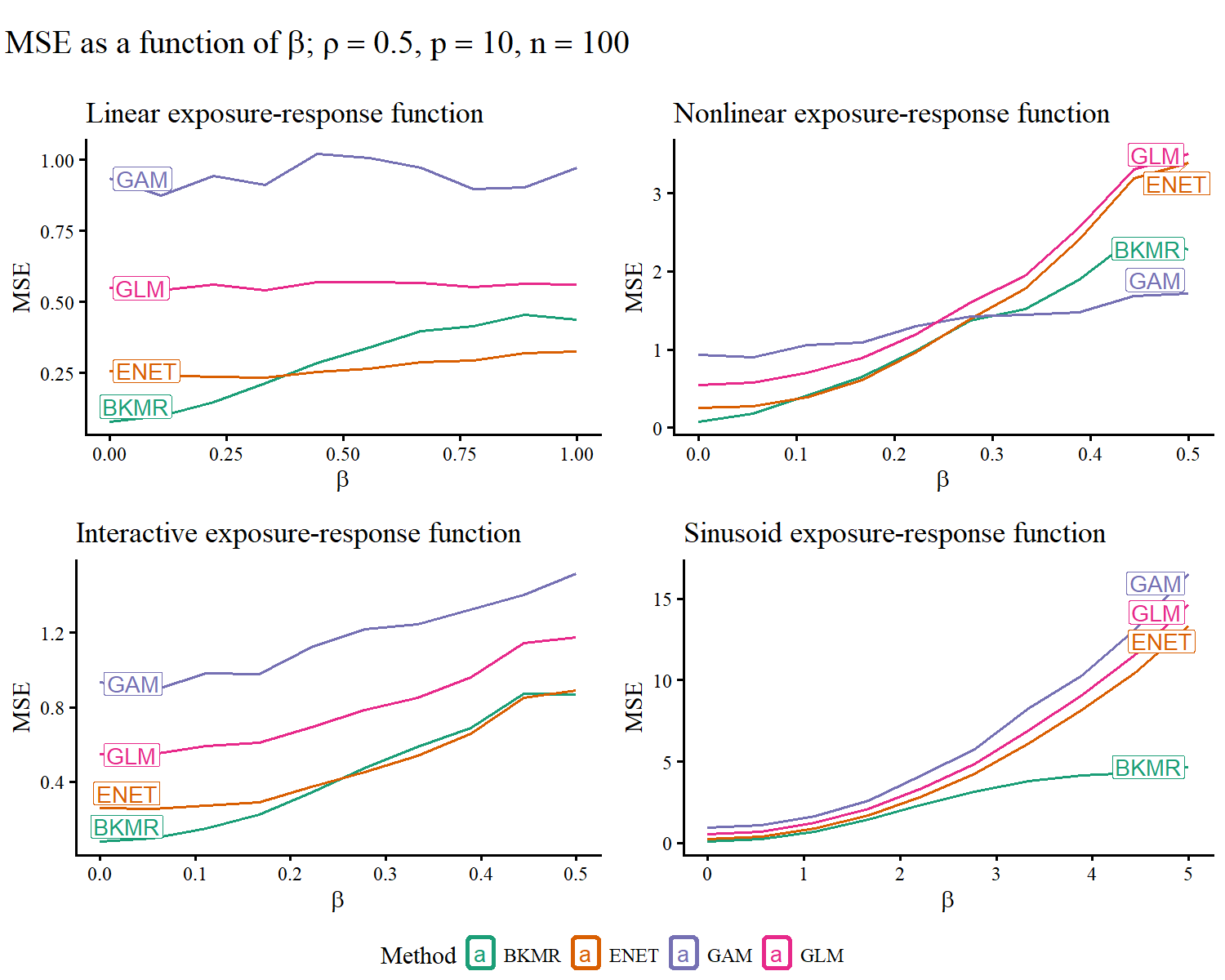}
    \caption{For each method (curve), the y-value is the prediction mean squared error (MSE) on new data, for different strengths of association $\beta$ (x-axis). Better methods have lower MSE curves. The new data are generated from the same exposure-response functions as the data used to fit the model, but have no random variation added. Exposure-response functions used are linear, nonlinear, linear interaction, and sinusoid (nonlinear interaction).}
    \label{fig:mse_n100_p10_rho5}
\end{figure}

\begin{figure}[H]
    \centering
    \includegraphics[width=0.9\linewidth]{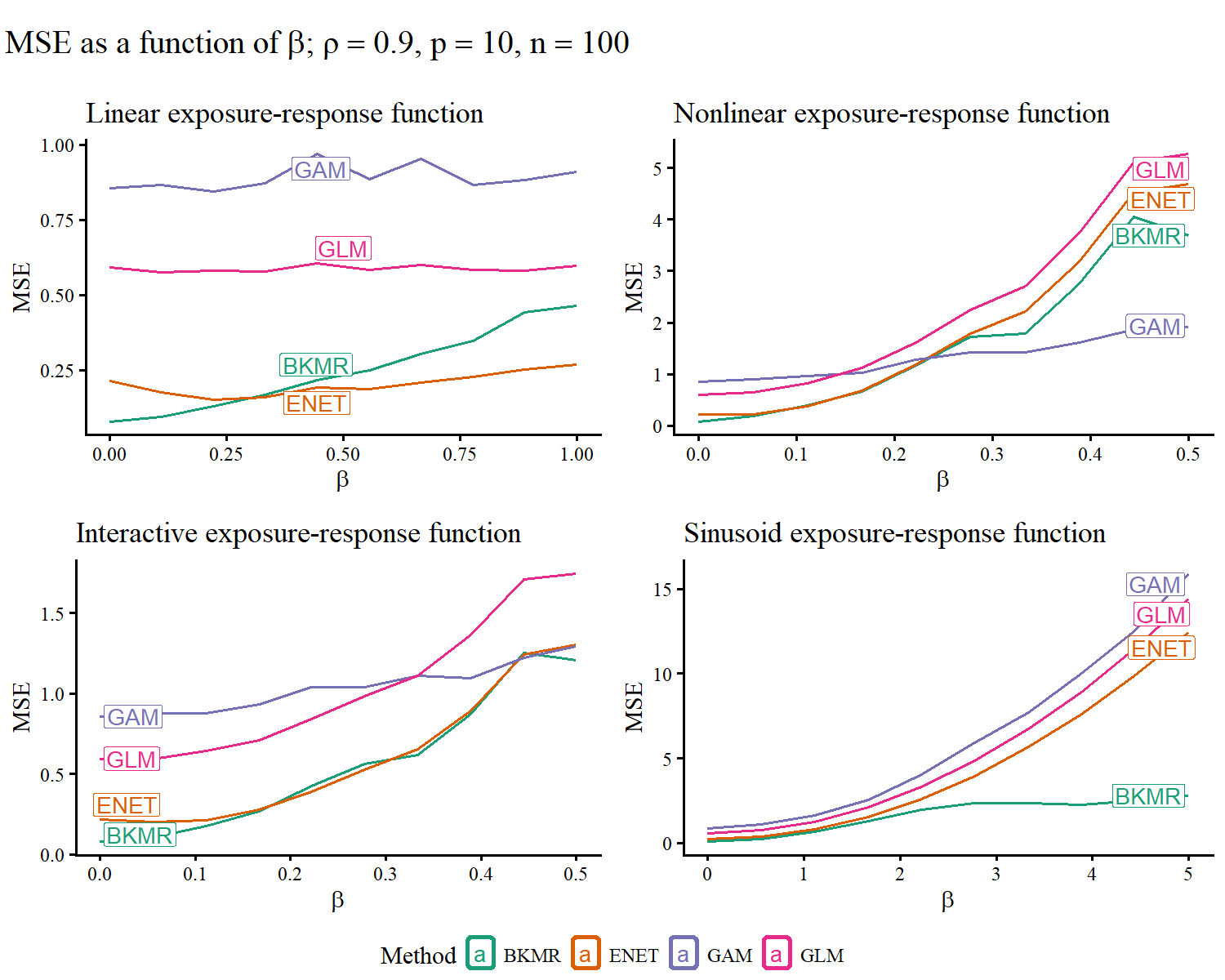}
    \caption{For each method (curve), the y-value is the prediction mean squared error (MSE) on new data, for different strengths of association $\beta$ (x-axis). Better methods have lower MSE curves. The new data are generated from the same exposure-response functions as the data used to fit the model, but have no random variation added. Exposure-response functions used are linear, nonlinear, linear interaction, and sinusoid (nonlinear interaction).}
    \label{fig:mse_n100_p10_rho9}
\end{figure}

\begin{figure}[H]
    \centering
    \includegraphics[width=0.9\linewidth]{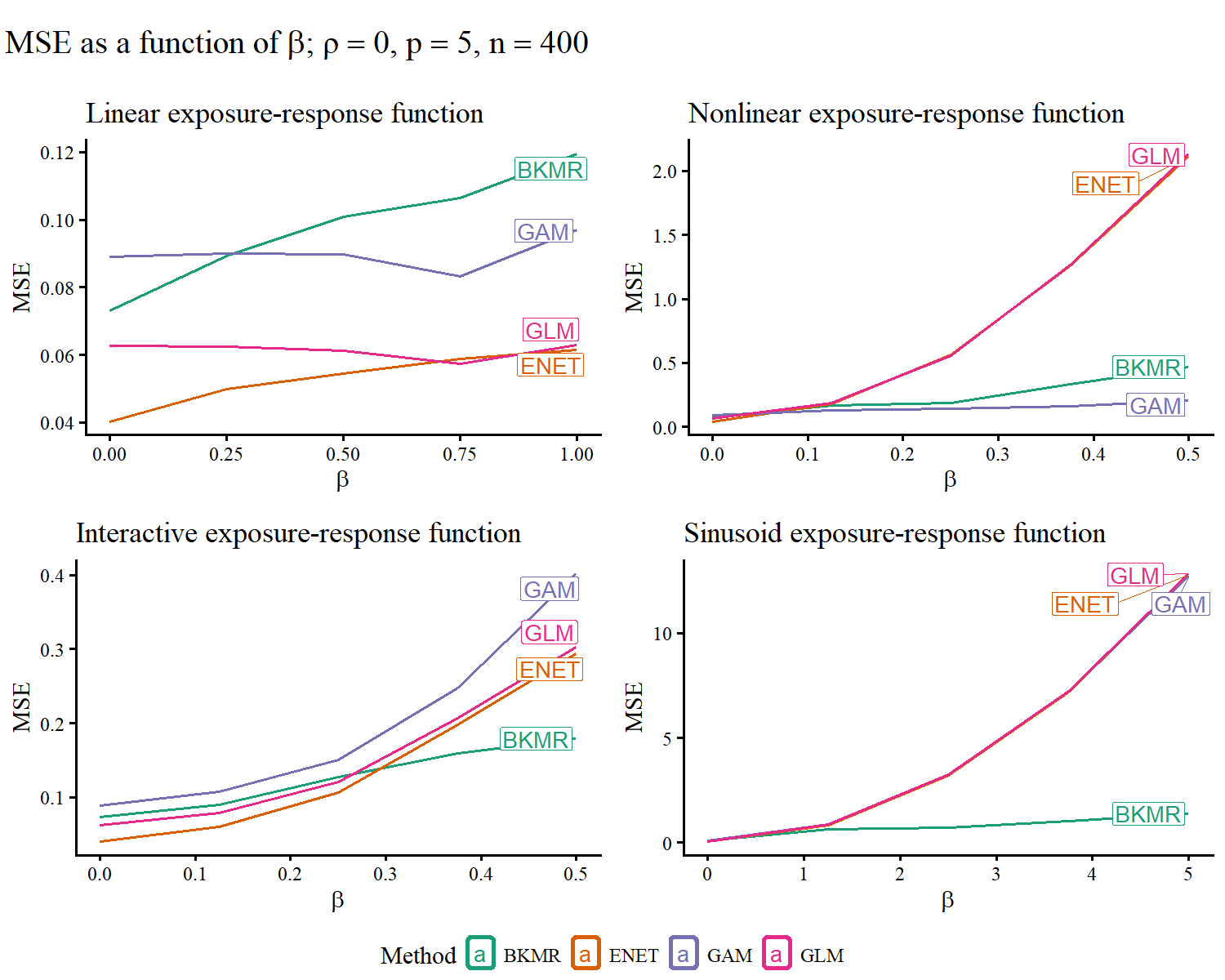}
    \caption{For each method (curve), the y-value is the prediction mean squared error (MSE) on new data, for different strengths of association $\beta$ (x-axis). Better methods have lower MSE curves. The new data are generated from the same exposure-response functions as the data used to fit the model, but have no random variation added. Exposure-response functions used are linear, nonlinear, linear interaction, and sinusoid (nonlinear interaction).}
    \label{fig:mse_n400_p5_rho0}
\end{figure}

\begin{figure}[H]
    \centering
    \includegraphics[width=0.9\linewidth]{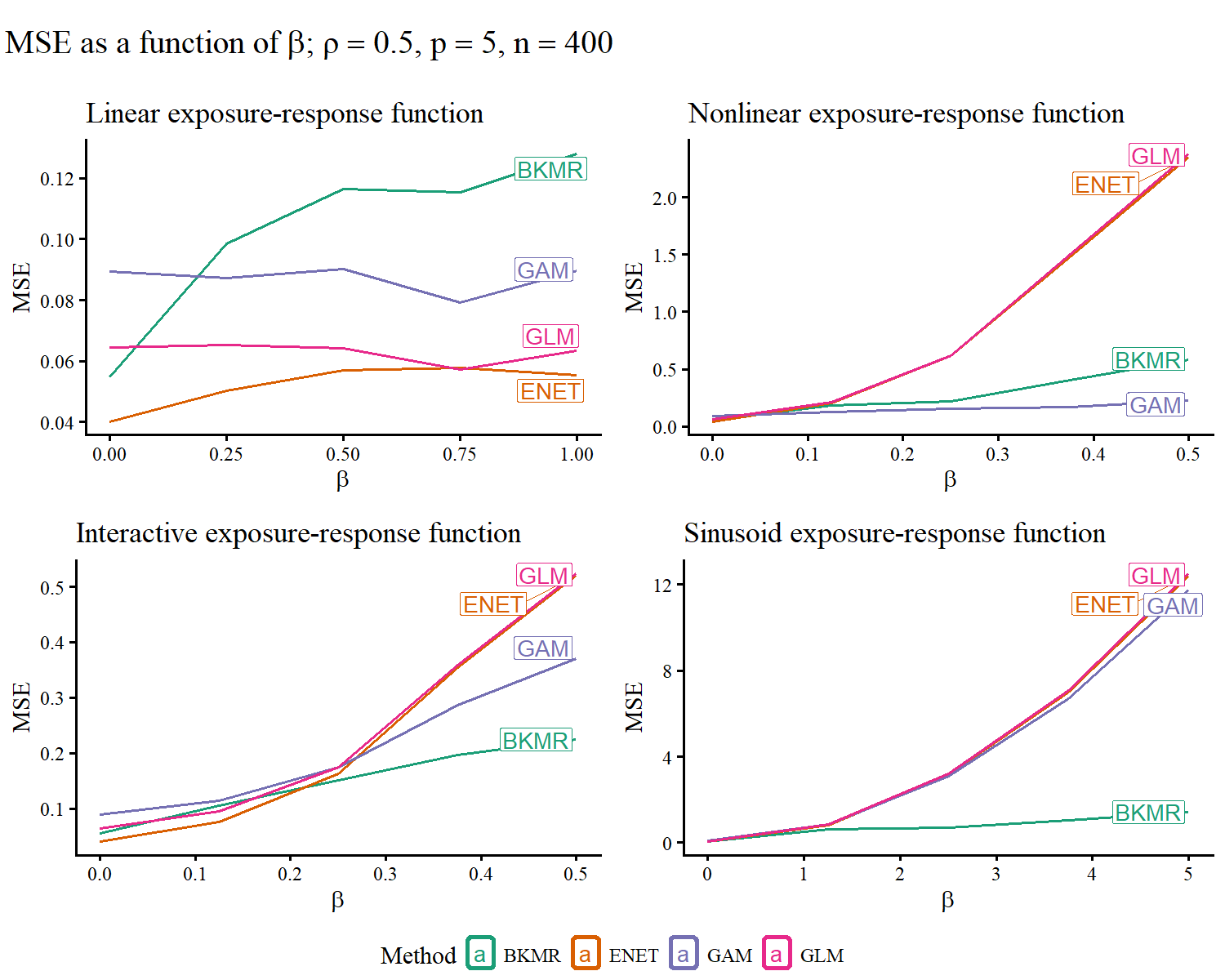}
    \caption{For each method (curve), the y-value is the prediction mean squared error (MSE) on new data, for different strengths of association $\beta$ (x-axis). Better methods have lower MSE curves. The new data are generated from the same exposure-response functions as the data used to fit the model, but have no random variation added. Exposure-response functions used are linear, nonlinear, linear interaction, and sinusoid (nonlinear interaction).}
    \label{fig:mse_n400_p5_rho5}
\end{figure}

\begin{figure}[H]
    \centering
    \includegraphics[width=0.5\linewidth]{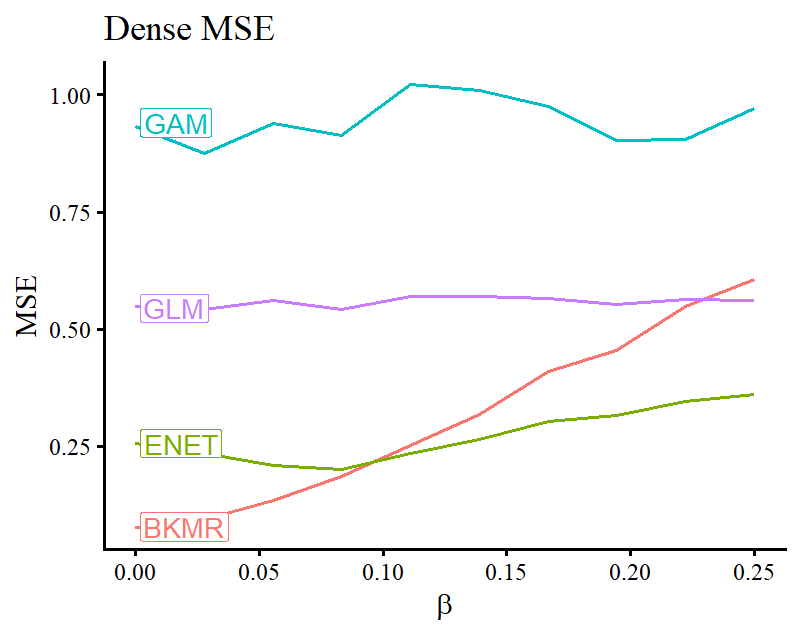}
    \caption{For each method (curve), the y-value is the prediction mean squared error (MSE) on new data, for different strengths of association $\beta$ (x-axis). Better methods have lower MSE curves. The new data are generated from the same exposure-response functions as the data used to fit the model, but have no random variation added. The exposure-response function used is dense (all exposures affecting the response) with $p=10$ exposures.}
    \label{fig:mse_n100_p10_rho5_dense}
\end{figure}

\begin{figure}[H]
    \centering
    \includegraphics[width=0.5\linewidth]{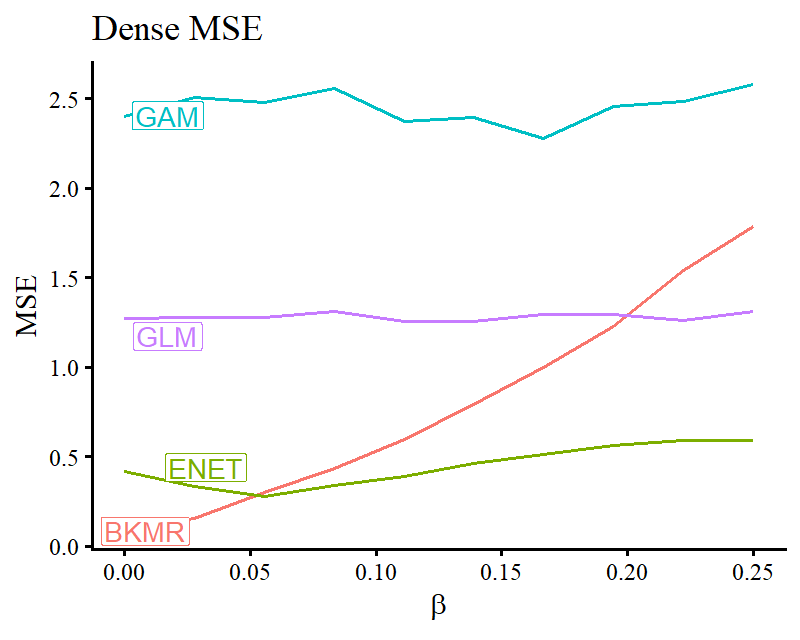}
    \caption{For each method (curve), the y-value is the prediction mean squared error (MSE) on new data, for different strengths of association $\beta$ (x-axis). Better methods have lower MSE curves. The new data are generated from the same exposure-response functions as the data used to fit the model, but have no random variation added.  The exposure-response function used is dense (all exposures affecting the response) with $p=20$ exposures.}
    \label{fig:mse_n100_p20_rho5_dense}
\end{figure}

\section{Results Tables}

\begin{figure}[H]
    \centering
    \includegraphics[width=0.7\linewidth]{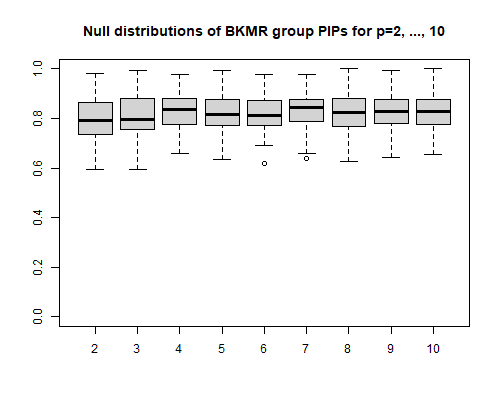}
    \caption{Null distribution of BKMR hierarchical variable selection group PIPs using package-default settings, with $n=100, \rho=0.5$ and 100 Monte Carlo replications. The number of exposure variables in the simulated data sets $p$ ranged from 2 to 10, all of which were included in a single group.}
    \label{fig:bkmr_null}
\end{figure}

This section provides results tables for all scenarios. The tables summarize the power curves, selecting representative values of $\beta$. The results tables display results for the value of $\beta$ which maximized the standard deviation of probability of rejecting $H_0$ in that scenario, among methods displayed in the table. BKMR with hierarchical variable selection's results are not displayed as the group PIP was essentially always greater than 0.5 in all scenarios tried. The null distribution of the group PIP for different values of $p$, with default tuning, is displayed in Figure \ref{fig:bkmr_null}.

\subsection{Individual component hypothesis tests results tables}

\begin{table}[H]
    \centering
    \small
\begin{tabular}{ll|llllllll}
\hline
   & $\beta$ & GLM & GAM (HT) & GAM (VS) & ENET & BKMR (0.50) & BKMR (0.95)\\
\hline
Lin. & 0.33 & 0.36 & 0.37 & 0.82 & 0.91 & 0.68 & 0.07 \\
N.L & 0.06 & 0.23 & 0.24 & 0.66 & 0.76 & 0.50 & 0.03 \\
Int. & 0.33 & 0.34 & 0.34 & 0.75 & 0.88 & 0.66 & 0.09 \\
Sine & 5.00 & 0.07 & 0.13 & 0.61 & 0.23 & 0.96 & 0.92 \\
None & 0.00 & 0.03 & 0.05 & 0.43 & 0.41 & 0.28 & 0.00 \\


\end{tabular}

    \caption{Individual component hypothesis test performance for $n=100, p=5, \rho=0$. Each row corresponds to a true exposure-response relationship (linear, nonlinear, linear interaction, sinusoidal, or none) along with a parameter $\beta$ that controls the strength of the exposure-response relationship, where $\beta=0$ corresponds to no relationship, and higher values correspond to stronger relationships. The columns for each method display the probability of rejecting the null hypothesis $H_0$: no association between treatment variable $a_1$ and the expected response $y$, estimated using Monte Carlo simulation. The targeted Type I error rate was $0.05$.}
    \label{tab:varsel_n100_p5_rho0}
\end{table}

\begin{table}[H]
    \centering
    \small
\begin{tabular}{ll|llllllll}
\hline
    & $\beta$ & GLM & GAM (HT) & GAM (VS) & ENET & BKMR (0.50) & BKMR (0.95)\\
\hline
Lin. & 0.44 & 0.42 & 0.44 & 0.85 & 0.92 & 0.81 & 0.14 \\
N.L & 0.11 & 0.38 & 0.40 & 0.76 & 0.84 & 0.69 & 0.17 \\
Int. & 0.22 & 0.26 & 0.30 & 0.72 & 0.83 & 0.65 & 0.09 \\
Sine & 5.00 & 0.09 & 0.28 & 0.70 & 0.16 & 0.99 & 0.95 \\
None & 0.00 & 0.05 & 0.07 & 0.41 & 0.38 & 0.30 & 0.00 \\


\end{tabular}

    \caption{Individual component hypothesis test performance for $n=100, p=5, \rho=0.5$. Each row corresponds to a true exposure-response relationship (linear, nonlinear, linear interaction, sinusoidal, or none) along with a parameter $\beta$ that controls the strength of the exposure-response relationship, where $\beta=0$ corresponds to no relationship, and higher values correspond to stronger relationships. The columns for each method display the probability of rejecting the null hypothesis $H_0$: no association between treatment variable $a_1$ and the expected response $y$, estimated using Monte Carlo simulation. The targeted Type I error rate was $0.05$.}
    \label{tab:varsel_n100_p5_rho5}
\end{table}

\begin{table}[H]
    \centering
    \small
\begin{tabular}{ll|llllll}
 & $\beta$ & GLM & GAM (HT) & GAM (VS) & ENET & BKMR (0.50) & BKMR (0.95)\\
\hline
Lin. & 0.67 & 0.20 & 0.21 & 0.71 & 0.88 & 0.70 & 0.05 \\
N.L & 0.22 & 0.35 & 0.39 & 0.83 & 0.90 & 0.80 & 0.17 \\
Int. & 0.33 & 0.19 & 0.24 & 0.74 & 0.83 & 0.73 & 0.06 \\
Sine & 3.33 & 0.05 & 0.19 & 0.57 & 0.18 & 0.98 & 0.86 \\
None & 0.00 & 0.06 & 0.07 & 0.42 & 0.33 & 0.36 & 0.00 \\


\end{tabular}

    \caption{Individual component hypothesis test performance for $n=100, p=5, \rho=0.9$. Each row corresponds to a true exposure-response relationship (linear, nonlinear, linear interaction, sinusoidal, or none) along with a parameter $\beta$ that controls the strength of the exposure-response relationship, where $\beta=0$ corresponds to no relationship, and higher values correspond to stronger relationships. The columns for each method display the probability of rejecting the null hypothesis $H_0$: no association between treatment variable $a_1$ and the expected response $y$, estimated using Monte Carlo simulation. The targeted Type I error rate was $0.05$.}
    \label{tab:varsel_n100_p5_rho9}
\end{table}

\begin{table}[H]
    \centering
    \small
\begin{tabular}{ll|lllllll}
\hline
   & $\beta$ & GLM & GAM (HT) & GAM (VS) & ENET & BKMR (0.50) & BKMR (0.95)\\
\hline
Lin. & 0.44 & 0.57 & 0.58 & 0.91 & 0.96 & 0.76 & 0.12 \\
N.L & 0.11 & 0.49 & 0.52 & 0.86 & 0.91 & 0.67 & 0.12 \\
Int. & 0.33 & 0.31 & 0.34 & 0.75 & 0.86 & 0.56 & 0.05 \\
Sine & 4.44 & 0.04 & 0.13 & 0.57 & 0.17 & 0.80 & 0.60 \\
None & 0.00 & 0.06 & 0.09 & 0.42 & 0.36 & 0.25 & 0.00 \\


\end{tabular}

    \caption{Individual component hypothesis test performance for $n=100, p=10, \rho=0$. Each row corresponds to a true exposure-response relationship (linear, nonlinear, linear interaction, sinusoidal, or none) along with a parameter $\beta$ that controls the strength of the exposure-response relationship, where $\beta=0$ corresponds to no relationship, and higher values correspond to stronger relationships. The columns for each method display the probability of rejecting the null hypothesis $H_0$: no association between treatment variable $a_1$ and the expected response $y$, estimated using Monte Carlo simulation. The targeted Type I error rate was $0.05$.}
    \label{tab:varsel_n100_p10_rho0}
\end{table}

\begin{table}[H]
    \centering
    \small
\begin{tabular}{ll|lllllll}
\hline
  & $\beta$ & GLM & GAM (HT) & GAM (VS) & ENET & BKMR (0.50) & BKMR (0.95)\\
\hline
Lin. & 0.44 & 0.35 & 0.39 & 0.80 & 0.88 & 0.68 & 0.08 \\
N.L & 0.11 & 0.30 & 0.34 & 0.74 & 0.84 & 0.56 & 0.06 \\
Int. & 0.33 & 0.34 & 0.32 & 0.76 & 0.89 & 0.64 & 0.09 \\
Sine & 5.00 & 0.07 & 0.26 & 0.62 & 0.15 & 0.96 & 0.88 \\
None & 0.00 & 0.04 & 0.06 & 0.40 & 0.32 & 0.25 & 0.00 \\


\end{tabular}

    \caption{Individual component hypothesis test performance for $n=100, p=10, \rho=0.5$. Each row corresponds to a true exposure-response relationship (linear, nonlinear, linear interaction, sinusoidal, or none) along with a parameter $\beta$ that controls the strength of the exposure-response relationship, where $\beta=0$ corresponds to no relationship, and higher values correspond to stronger relationships. The columns for each method display the probability of rejecting the null hypothesis $H_0$: no association between treatment variable $a_1$ and the expected response $y$, estimated using Monte Carlo simulation. The targeted Type I error rate was $0.05$.}
    \label{tab:varsel_n100_p10_rho5}
\end{table}

\begin{table}[H]
    \centering
    \small
\begin{tabular}{ll|lllllll}
\hline
  & $\beta$ & GLM & GAM (HT) & GAM (VS) & ENET & BKMR (0.50) & BKMR (0.95)\\
\hline
Lin. & 0.89 & 0.27 & 0.27 & 0.77 & 0.93 & 0.55 & 0.03 \\
N.L & 0.28 & 0.37 & 0.41 & 0.82 & 0.90 & 0.66 & 0.12 \\
Int. & 0.44 & 0.25 & 0.26 & 0.69 & 0.82 & 0.55 & 0.04 \\
Sine & 5.00 & 0.06 & 0.20 & 0.56 & 0.12 & 1.00 & 1.00 \\
None & 0.00 & 0.04 & 0.05 & 0.40 & 0.24 & 0.22 & 0.00 \\

\end{tabular}

    \caption{Individual component hypothesis test performance for $n=100, p=10, \rho=0.9$. Each row corresponds to a true exposure-response relationship (linear, nonlinear, linear interaction, sinusoidal, or none) along with a parameter $\beta$ that controls the strength of the exposure-response relationship, where $\beta=0$ corresponds to no relationship, and higher values correspond to stronger relationships. The columns for each method display the probability of rejecting the null hypothesis $H_0$: no association between treatment variable $a_1$ and the expected response $y$, estimated using Monte Carlo simulation. The targeted Type I error rate was $0.05$.}
    \label{tab:varsel_n100_p10_rho9}
\end{table}

\begin{table}[H]
    \centering
    \small
\begin{tabular}{ll|lllllll}
\hline
   & $\beta$ & GLM & GAM (HT) & GAM (VS) & ENET & BKMR (0.50) & BKMR (0.95)\\
\hline
Lin. & 0.25 & 0.69 & 0.69 & 0.93 & 0.96 & 0.82 & 0.22 \\
N.L & 0.12 & 0.96 & 0.98 & 1.00 & 1.00 & 0.98 & 0.87 \\
Int. & 0.12 & 0.22 & 0.21 & 0.69 & 0.79 & 0.50 & 0.04 \\
Sine & 5.00 & 0.13 & 0.43 & 0.81 & 0.06 & 1.00 & 1.00 \\
None & 0.00 & 0.06 & 0.06 & 0.48 & 0.34 & 0.30 & 0.00 \\


\end{tabular}

    \caption{Individual component hypothesis test performance for $n=400, p=5, \rho=0$. Each row corresponds to a true exposure-response relationship (linear, nonlinear, linear interaction, sinusoidal, or none) along with a parameter $\beta$ that controls the strength of the exposure-response relationship, where $\beta=0$ corresponds to no relationship, and higher values correspond to stronger relationships. The columns for each method display the probability of rejecting the null hypothesis $H_0$: no association between treatment variable $a_1$ and the expected response $y$, estimated using Monte Carlo simulation. The targeted Type I error rate was $0.05$.}
    \label{tab:varsel_n400_p5_rho0}
\end{table}

\begin{table}[H]
    \centering
    \small
\begin{tabular}{ll|lllllll}
\hline
  & $\beta$ & GLM & GAM (HT) & GAM (VS) & ENET & BKMR (0.50) & BKMR (0.95)\\
\hline
Lin. & 0.25 & 0.53 & 0.54 & 0.88 & 0.95 & 0.80 & 0.14 \\
N.L & 0.12 & 0.94 & 0.95 & 0.99 & 1.00 & 0.98 & 0.81 \\
Int. & 0.12 & 0.34 & 0.34 & 0.82 & 0.89 & 0.69 & 0.09 \\
Sine & 5.00 & 0.18 & 0.61 & 0.92 & 0.03 & 1.00 & 1.00 \\
None & 0.00 & 0.06 & 0.07 & 0.43 & 0.36 & 0.42 & 0.00 \\


\end{tabular}

    \caption{Individual component hypothesis test performance for $n=400, p=4, \rho=0.5$. Each row corresponds to a true exposure-response relationship (linear, nonlinear, linear interaction, sinusoidal, or none) along with a parameter $\beta$ that controls the strength of the exposure-response relationship, where $\beta=0$ corresponds to no relationship, and higher values correspond to stronger relationships. The columns for each method display the probability of rejecting the null hypothesis $H_0$: no association between treatment variable $a_1$ and the expected response $y$, estimated using Monte Carlo simulation. The targeted Type I error rate was $0.05$.}
    \label{tab:varsel_n400_p5_rho5}
\end{table}

\subsection{Whole-mixture hypothesis test results tables}

\begin{table}[H]
    \centering
    \small
\begin{tabular}{{p{0.04\linewidth}p{0.04\linewidth}|p{0.11\linewidth}p{0.11\linewidth}p{0.11\linewidth}p{0.11\linewidth}p{0.12\linewidth}p{0.11\linewidth}p{0.05\linewidth}}}
\hline
  &  $\beta$ & GLM & GAM (HT, bonf.) & PCR&  BKMR (Contrast) & BKMR (Hier., 0.95) & WQS (Pos.) & QGC \\
\hline
Lin. & 0.56 & 0.82 & 0.76 & 0.79 & 0.28 & 0.70 & 0.47 & 0.47 \\
N.L. & 0.17 & 0.88 & 0.85 & 0.84 & 0.16 & 0.84 & 0.35 & 0.34 \\
Int. & 0.50 & 0.69 & 0.58 & 0.66 & 0.18 & 0.56 & 0.39 & 0.41 \\
Sine & 5.00 & 0.09 & 0.14 & 0.09 & 0.42 & 0.14 & 0.09 & 0.14 \\
None & 0.00 & 0.06 & 0.06 & 0.07 & 0.00 & 0.04 & 0.05 & 0.06 \\


\end{tabular}

    \caption{Whole-mixture performance for $n=100, p=5, \rho=0$. Each row corresponds to a true exposure-response relationship (linear, nonlinear, linear interaction, sinusoidal, or none) along with a parameter $\beta$ that controls the strength of the exposure-response relationship, where $\beta=0$ corresponds to no relationship, and higher values correspond to stronger relationships. The columns for each method display the probability of rejecting the null hypothesis $H_0$: no association between any treatment variables $a_1, ..., a_p$ and the expected response $y$, estimated using Monte Carlo simulation. The targeted Type I error rate was $0.05$.}
    \label{tab:index_n100_p5_rho0}
\end{table}

\begin{table}[H]
    \centering
    \small
\begin{tabular}{{p{0.04\linewidth}p{0.04\linewidth}|p{0.11\linewidth}p{0.11\linewidth}p{0.11\linewidth}p{0.11\linewidth}p{0.12\linewidth}p{0.11\linewidth}p{0.05\linewidth}}}

\hline
 &  $\beta$ & GLM & GAM (HT, bonf.) & PCR&  BKMR (Contrast) & BKMR (Hier., 0.95) & WQS (Pos.) & QGC \\
\hline
Lin. & 0.44 & 0.80 & 0.37 & 0.84 & 0.41 & 0.84 & 0.59 & 0.79 \\
N.L. & 0.17 & 0.94 & 0.70 & 0.95 & 0.35 & 0.96 & 0.70 & 0.84 \\
Int. & 0.39 & 0.87 & 0.55 & 0.92 & 0.29 & 0.90 & 0.56 & 0.79 \\
Sine & 5.00 & 0.14 & 0.24 & 0.16 & 0.56 & 0.34 & 0.22 & 0.43 \\
None & 0.00 & 0.05 & 0.09 & 0.04 & 0.00 & 0.03 & 0.04 & 0.04 \\


\end{tabular}

    \caption{Whole-mixture performance for $n=100, p=5, \rho=0.5$. Each row corresponds to a true exposure-response relationship (linear, nonlinear, linear interaction, sinusoidal, or none) along with a parameter $\beta$ that controls the strength of the exposure-response relationship, where $\beta=0$ corresponds to no relationship, and higher values correspond to stronger relationships. The columns for each method display the probability of rejecting the null hypothesis $H_0$: no association between any treatment variables $a_1, ..., a_p$ and the expected response $y$, estimated using Monte Carlo simulation. The targeted Type I error rate was $0.05$.}
    \label{tab:index_n100_p5_rho5}
\end{table}

\begin{table}[H]
    \centering
    \small
\begin{tabular}{{p{0.04\linewidth}p{0.04\linewidth}|p{0.11\linewidth}p{0.11\linewidth}p{0.11\linewidth}p{0.11\linewidth}p{0.12\linewidth}p{0.11\linewidth}p{0.05\linewidth}}}

\hline
  &  $\beta$ & GLM & GAM (HT, bonf.) & PCR&  BKMR (Contrast) & BKMR (Hier., 0.95) & WQS (Pos.) & QGC \\
\hline
Lin. & 0.44 & 0.89 & 0.14 & 0.97 & 0.61 & 0.97 & 0.78 & 0.92 \\
N.L. & 0.11 & 0.74 & 0.20 & 0.91 & 0.20 & 0.88 & 0.48 & 0.67 \\
Int. & 0.28 & 0.86 & 0.22 & 0.96 & 0.26 & 0.94 & 0.58 & 0.81 \\
Sine & 5.00 & 0.26 & 0.31 & 0.35 & 0.70 & 0.85 & 0.43 & 0.78 \\
None & 0.00 & 0.03 & 0.08 & 0.05 & 0.00 & 0.03 & 0.03 & 0.04 \\

\end{tabular}

    \caption{Whole-mixture performance for $n=100, p=5, \rho=0.9$. Each row corresponds to a true exposure-response relationship (linear, nonlinear, linear interaction, sinusoidal, or none) along with a parameter $\beta$ that controls the strength of the exposure-response relationship, where $\beta=0$ corresponds to no relationship, and higher values correspond to stronger relationships. The columns for each method display the probability of rejecting the null hypothesis $H_0$: no association between any treatment variables $a_1, ..., a_p$ and the expected response $y$, estimated using Monte Carlo simulation. The targeted Type I error rate was $0.05$.}
    \label{tab:index_n100_p5_rho9}
\end{table}

\begin{table}[H]
    \centering
    \small
\begin{tabular}{{p{0.04\linewidth}p{0.04\linewidth}|p{0.11\linewidth}p{0.11\linewidth}p{0.11\linewidth}p{0.11\linewidth}p{0.12\linewidth}p{0.11\linewidth}p{0.05\linewidth}}}

\hline
   &  $\beta$ & GLM & GAM (HT, bonf.) & PCR&  BKMR (Contrast) & BKMR (Hier., 0.95) & WQS (Pos.) & QGC \\
\hline
Lin. & 0.67 & 0.85 & 0.86 & 0.81 & 0.24 & 0.80 & 0.52 & 0.37 \\
N.L. & 0.17 & 0.80 & 0.80 & 0.76 & 0.08 & 0.76 & 0.27 & 0.20 \\
Int. & 0.50 & 0.56 & 0.55 & 0.52 & 0.06 & 0.48 & 0.32 & 0.22 \\
Sine & 5.00 & 0.05 & 0.13 & 0.06 & 0.24 & 0.10 & 0.08 & 0.11 \\
None & 0.00 & 0.03 & 0.06 & 0.04 & 0.00 & 0.02 & 0.07 & 0.04 \\


\end{tabular}

    \caption{Whole-mixture performance for $n=100, p=10, \rho=0$. Each row corresponds to a true exposure-response relationship (linear, nonlinear, linear interaction, sinusoidal, or none) along with a parameter $\beta$ that controls the strength of the exposure-response relationship, where $\beta=0$ corresponds to no relationship, and higher values correspond to stronger relationships. The columns for each method display the probability of rejecting the null hypothesis $H_0$: no association between any treatment variables $a_1, ..., a_p$ and the expected response $y$, estimated using Monte Carlo simulation. The targeted Type I error rate was $0.05$.}
    \label{tab:index_n100_p10_rho0}
\end{table}

\begin{table}[H]
    \centering
    \small
\begin{tabular}{{p{0.04\linewidth}p{0.05\linewidth}|p{0.11\linewidth}p{0.11\linewidth}p{0.11\linewidth}p{0.11\linewidth}p{0.12\linewidth}p{0.11\linewidth}p{0.05\linewidth}}}

\hline
  &  $\beta$ & GLM & GAM (HT, bonf.) & PCR&  BKMR (Contrast) & BKMR (Hier., 0.95) & WQS (Pos.) & QGC \\
\hline
Lin. & 0.44 & 0.70 & 0.26 & 0.76 & 0.37 & 0.83 & 0.62 & 0.80 \\
N.L. & 0.17 & 0.87 & 0.54 & 0.90 & 0.30 & 0.93 & 0.64 & 0.83 \\
Int. & 0.39 & 0.82 & 0.38 & 0.86 & 0.23 & 0.89 & 0.62 & 0.80 \\
Sine & 5.00 & 0.10 & 0.20 & 0.12 & 0.44 & 0.30 & 0.22 & 0.42 \\
Opp. & -0.78 & 0.88 & 0.81 & 0.68 & 0.04 & 0.58 & 0.22 & 0.14 \\
None & 0.00 & 0.04 & 0.08 & 0.03 & 0.00 & 0.01 & 0.05 & 0.06 \\


\end{tabular}

    \caption{Whole-mixture performance for $n=100, p=10, \rho=0.5$. Each row corresponds to a true exposure-response relationship (linear, nonlinear, linear interaction, sinusoidal, opposite, or none) along with a parameter $\beta$ that controls the strength of the exposure-response relationship, where $\beta=0$ corresponds to no relationship, and higher values correspond to stronger relationships, except for the opposite-effects scenario. In the opposite effects scenario, $E(y|a_1, a_2) = \beta a_1 + a_2$. The columns for each method display the probability of rejecting the null hypothesis $H_0$: no association between any treatment variables $a_1, ..., a_p$ and the expected response $y$, estimated using Monte Carlo simulation. The targeted Type I error rate was $0.05$.}
    \label{tab:index_n100_p10_rho5}
\end{table}
    
\begin{table}[H]
    \centering
    \small
\begin{tabular}{{p{0.04\linewidth}p{0.04\linewidth}|p{0.11\linewidth}p{0.11\linewidth}p{0.11\linewidth}p{0.11\linewidth}p{0.12\linewidth}p{0.11\linewidth}p{0.05\linewidth}}}

\hline
  &  $\beta$ & GLM & GAM (HT, bonf.) & PCR&  BKMR (Contrast) & BKMR (Hier., 0.95) & WQS (Pos.) & QGC \\
\hline
Lin. & 0.67 & 0.98 & 0.09 & 1.00 & 0.92 & 1.00 & 0.96 & 1.00 \\
N.L. & 0.17 & 0.91 & 0.16 & 0.99 & 0.35 & 0.99 & 0.77 & 0.92 \\
Int. & 0.33 & 0.90 & 0.14 & 0.98 & 0.29 & 0.97 & 0.72 & 0.86 \\
Sine & 5.00 & 0.17 & 0.25 & 0.38 & 0.67 & 0.82 & 0.49 & 0.74 \\
None & 0.00 & 0.04 & 0.05 & 0.05 & 0.01 & 0.04 & 0.06 & 0.05 \\


\end{tabular}

    \caption{Whole-mixture performance for $n=100, p=10, \rho=0.9$. Each row corresponds to a true exposure-response relationship (linear, nonlinear, linear interaction, sinusoidal, or none) along with a parameter $\beta$ that controls the strength of the exposure-response relationship, where $\beta=0$ corresponds to no relationship, and higher values correspond to stronger relationships. The columns for each method display the probability of rejecting the null hypothesis $H_0$: no association between any treatment variables $a_1, ..., a_p$ and the expected response $y$, estimated using Monte Carlo simulation. The targeted Type I error rate was $0.05$.}
    \label{tab:index_n100_p10_rho9}
\end{table}

\begin{table}[H]
    \centering
    \small
\begin{tabular}{{p{0.04\linewidth}p{0.04\linewidth}|p{0.11\linewidth}p{0.11\linewidth}p{0.11\linewidth}p{0.11\linewidth}p{0.12\linewidth}p{0.11\linewidth}p{0.05\linewidth}}}

\hline
  &  $\beta$ & GLM & GAM (HT, bonf.) & PCR&  BKMR (Contrast) & BKMR (Hier., 0.95) & WQS (Pos.) & QGC \\
\hline
Lin. & 0.25 & 0.76 & 0.68 & 0.74 & 0.20 & 0.56 & 0.48 & 0.44 \\
N.L. & 0.12 & 1.00 & 1.00 & 0.99 & 0.48 & 1.00 & 0.76 & 0.66 \\
Int. & 0.25 & 0.76 & 0.64 & 0.70 & 0.16 & 0.52 & 0.41 & 0.47 \\
Sine & 3.75 & 0.10 & 0.31 & 0.09 & 0.95 & 0.99 & 0.11 & 0.36 \\
None & 0.00 & 0.04 & 0.06 & 0.06 & 0.00 & 0.02 & 0.04 & 0.06 \\


\end{tabular}

    \caption{Whole-mixture performance for $n=400, p=5, \rho=0$. Each row corresponds to a true exposure-response relationship (linear, nonlinear, linear interaction, sinusoidal, or none) along with a parameter $\beta$ that controls the strength of the exposure-response relationship, where $\beta=0$ corresponds to no relationship, and higher values correspond to stronger relationships. The columns for each method display the probability of rejecting the null hypothesis $H_0$: no association between any treatment variables $a_1, ..., a_p$ and the expected response $y$, estimated using Monte Carlo simulation. The targeted Type I error rate was $0.05$.}
    \label{tab:index_n400_p5_rho0}
\end{table}

\begin{table}[H]
    \centering
    \small
\begin{tabular}{{p{0.04\linewidth}p{0.04\linewidth}|p{0.04\linewidth}p{0.11\linewidth}p{0.11\linewidth}p{0.11\linewidth}p{0.12\linewidth}p{0.11\linewidth}p{0.05\linewidth}}}

\hline
  &  $\beta$ & GLM & GAM (HT, bonf.) & PCR&  BKMR (Contrast) & BKMR (Hier., 0.95) & WQS (Pos.) & QGC \\
\hline
Lin. & 0.25 & 0.90 & 0.44 & 0.89 & 0.50 & 0.89 & 0.75 & 0.88 \\
N.L. & 0.12 & 1.00 & 0.98 & 1.00 & 0.58 & 1.00 & 0.92 & 1.00 \\
Int. & 0.12 & 0.66 & 0.28 & 0.68 & 0.10 & 0.62 & 0.38 & 0.52 \\
Sine & 3.75 & 0.34 & 0.68 & 0.41 & 0.96 & 0.99 & 0.48 & 0.93 \\
None & 0.00 & 0.06 & 0.06 & 0.06 & 0.00 & 0.04 & 0.10 & 0.06 \\


\end{tabular}

    \caption{Whole-mixture performance for $n=400, p=5, \rho=0.5$. Each row corresponds to a true exposure-response relationship (linear, nonlinear, linear interaction, sinusoidal, or none) along with a parameter $\beta$ that controls the strength of the exposure-response relationship, where $\beta=0$ corresponds to no relationship, and higher values correspond to stronger relationships. The columns for each method display the probability of rejecting the null hypothesis $H_0$: no association between any treatment variables $a_1, ..., a_p$ and the expected response $y$, estimated using Monte Carlo simulation. The targeted Type I error rate was $0.05$.}
    \label{tab:index_n400_p5_rho5}
\end{table}

\begin{table}[H]
    \centering
    \small
\begin{tabular}{{p{0.04\linewidth}p{0.04\linewidth}p{0.04\linewidth}|p{0.04\linewidth}p{0.11\linewidth}p{0.04\linewidth}p{0.11\linewidth}p{0.12\linewidth}p{0.11\linewidth}p{0.05\linewidth}}}

\hline
  & $p$ &  $\beta$ & GLM & GAM (HT, bonf.) & PCR & BKMR (Contrast) & BKMR (Hier., 0.95) &  WQS (Pos.) & QGC \\
\hline
Lin. &  10 & 0.17 & 0.96 & 0.16 & 0.99 & 0.89 & 1.00 & 0.96 & 1.00 \\
Lin. &  20 & 0.08 & 0.89 & 0.14 & 0.99 & 0.89 & 1.00 & 0.95 & 1.00 \\
None & 10 & 0.00 & 0.04 & 0.08 & 0.03 & 0.00 & 0.01 & 0.05 & 0.06 \\
None & 20 & 0.00 & 0.08 & 0.14 & 0.07 & 0.00 & 0.04 & 0.05 & 0.08 \\
\end{tabular}
    \caption{Whole-mixture performance for the dense scenarios with $n=100, \rho=0.5$. Each row corresponds to a true exposure-response relationship (linear  or none) along with a parameter $\beta$ that controls the strength of the exposure-response relationship, where $\beta=0$ corresponds to no relationship, and higher values correspond to stronger relationships. The columns for each method display the probability of rejecting the null hypothesis $H_0$: no association between any treatment variables $a_1, ..., a_p$ and the expected response $y$, estimated using Monte Carlo simulation. The targeted Type I error rate was $0.05$.}
    \label{tab:index_dense}
\end{table}

\section{Other supporting information}

\begin{singlespace}
    \printbibliography
\end{singlespace}

%% file: commands.tex
\newcommand{\ba}{\mbox{\bf a}}

\newcommand{\bx}{\mbox{\bf x}}

\newcommand{\beq}{ \begin{equation}}
\newcommand{\eeq}{ \end{equation}}
\newcommand{\beqn}{ \begin{eqnarray}}
\newcommand{\eeqn}{ \end{eqnarray}}

\usepackage{caption}